%% file: review.tex
\documentclass[final 3p,times,pdflatex]{elsarticle}

\journal{Computer Physics Communications}

\usepackage{graphicx}
\usepackage[a4paper,margin=2.4cm,bottom=3cm]{geometry}

\usepackage{booktabs}
\usepackage{multirow}
\usepackage{mathtools}

\usepackage{amsmath}
\usepackage{nicefrac}

\usepackage[T1]{fontenc}
\usepackage[utf8]{inputenc}       
\usepackage{anyfontsize}
\usepackage{amssymb}
\usepackage{slashed}              
\usepackage{dsfont}               
\usepackage{mathrsfs}             
\usepackage{amsfonts}

\usepackage{xspace}

\biboptions{comma,square,sort&compress} 

\usepackage[figuresright]{rotating}

\usepackage[usenames]{color}
\usepackage{colortbl}

\definecolor{violet}{RGB}{111,0,255}
\definecolor{webgreen}{rgb}{0,0.75,0}
\definecolor{webred}{rgb}{0.75,0,0}
\definecolor{webblue}{rgb}{0,0,0.75}
\definecolor{darkblue}{rgb}{0,0,0.6}
\definecolor{darkgreen}{rgb}{0,0.5,0.5}
\definecolor{darkpurple}{rgb}{0.5,0,0.5}
\definecolor{darkorange}{rgb}{1,0.5,0}
\definecolor{darkgrey}{rgb}{0.4,0.4,0.4}
\definecolor{lgray}{rgb}{0.95,0.95,0.95}
\definecolor{lgreen}{rgb}{0.95,1.00,0.90}
\definecolor{lred}{rgb}{1.00,0.90,0.80}
\definecolor{lblue}{rgb}{0.2,0.35,1.00}
\definecolor{shadecolor}{rgb}{1.00,0.92,0.82}

\newcommand{\beq}{\begin{equation}}
\newcommand{\eeq}{\end{equation}}

\newcommand{\pslash}{\slashed{p}}
\newcommand{\qslash}{\slashed{q}}

\newcommand{\lb}{\left(}
\newcommand{\rb}{\right)}
\newcommand{\bracket}[1]{\left( #1 \right)}
\newcommand{\lsb}{\left[}
\newcommand{\rsb}{\right]}

\newcommand{\ii}{\mathrm{i}}

\begin{document}

\begin{frontmatter}

\title{Recent developments in bound-state calculations using the Dyson-Schwinger and Bethe-Salpeter equations}
\author[Graz]{H\`elios Sanchis-Alepuz}
\ead{helios.sanchis-alepuz@uni-graz.at}
\author[Giessen]{Richard Williams\corref{cor1}}
\ead{richard.williams@theo.physik.uni-giessen.de}
\cortext[cor1]{Corresponding author}

\address[Graz]{Institute of Physics, NAWI Graz, University of Graz,
Universit\"atsplatz 5, 8010 Graz, Austria}
\address[Giessen]{Institut f\"ur Theoretische Physik, Justus-Liebig--Universit\"at Giessen, 35392 Giessen, Germany}

\begin{abstract}
We review in detail modern numerical methods used in the determination and solution of Bethe-Salpeter and Dyson--Schwinger equations. The algorithms and techniques described are applicable to both the rainbow-ladder truncation and its non-trivial extensions. We discuss pedagogically the steps involved in constructing conventional mesons and baryons as systems of two- and three-quarks respectively.
As further application we describe the self-consistent calculation of form-factors and highlight the challenges that remain therein.
\end{abstract}

\begin{keyword}
Dyson-Schwinger approach
\sep
Bethe-Salpeter/Faddeev Equations
\sep
Non-perturbative methods
\sep
Hadrons
\end{keyword}

\end{frontmatter}

\newpage\input{1.introduction}
\newpage\input{2.quarks}

\newpage\input{3.mesons}
\newpage\input{4.baryons}
\newpage\input{5.formfactors}
\newpage\input{6.conclusions}
\appendix
\newpage\input{7.appendix}

\newpage\clearpage
\bibliographystyle{utphys}
\bibliography{review2}

\end{document}

%% file: 1.introduction.tex
\section{Introduction}\label{sec:introduction}
In functional approaches to Quantum Field Theories (QFTs), the Dyson-Schwinger equations~\cite{Dyson:1949ha,Schwinger:1951ex,Schwinger:1951hq} (DSEs) can be succinctly thought of as the equations of motion of Green's functions. They are exact---in principle---and are derived by considering the Ward identity associated with translational invariance. The dynamical information contained within these Green's functions pertains to the fundamental degrees of freedom, with the simplest being that of two-point functions that encode the propagation of a particle from one space-time point to another. When subject to a Fourier transformation, they may be recast into functions of momentum which thence describe the particles behaviour as a function of the energy scale. Higher $n$-point functions provide access to interaction vertices, together with more complicated constructs from which one may extract---depending upon the underlying theory---bound-states and scattering amplitudes.

Such extraction can be somewhat simplified when we all that is needed is the on-shell information about the bound-states of the theory. In that case, a different integral representation of the $n$-point functions may be used: the Dyson series. In this paper we will focus upon the theory of the strong interaction, Quantum Chromodynamics (QCD), and the calculation of both Green's functions and bound-states therein. Here, the bound-states (i.e. the hadrons) appear as poles in the relevant Green's function for select momentum configurations. For example, mesons appear as poles in a Green's function with four quark-legs (i.e. 4-point quark Green's function). Expansion around such a pole reveals the simpler Bethe-Salpeter equation (BSE) for the bound-state. This equation still depends upon the propagation and interaction of the constituents of the bound-state, therefore a combination of DSE and BSE methods is necessary. Discussions of these techniques can be found in various reviews, see 
e.g.~\cite{Roberts:1994dr,Alkofer:2000wg,Maris:2003vk,Fischer:2006ub,Binosi:2009qm,Swanson:2010pw,Bashir:2012fs,Eichmann:2016yit}.

In general, the solution of both the DSEs and the BSEs is a very complicated task, explaining why their introduction as non-perturbative tools in QFTs was initially restricted to formal applications. With advances in computing power, together with the development of ever sophisticated tools for performing algebraic manipulation~\cite{Mertig:1990an,Vermaseren:2000nd,Alkofer:2008nt,Huber:2011xc,Huber:2011qr,Cyrol:2016zqb,Shtabovenko:2016sxi}, robust treatments of DSEs~\cite{Maris:1994ux,Bloch:1995dd,Hauck:1996sm,McKay:1996th,Maris:1996zg,Atkinson:1997tu,Hauck:1998fz,Fischer:2003rp,Maas:2005xh} and BSEs~\cite{Munczek:1991jb,Jain:1991pk,Jain:1993qh,Maris:1997tm,Oettel:2001kd} appeared whose initial numerical techniques underpin much of what is done today. Since then, their use has blossomed especially in the context of theoretical hadron physics. However, until very recently the computer capabilities and available numerical tools has restricted their use to very simplified setups which, in turn, has limited the value of DSE/BSE calculations to qualitative rather than quantitative calculations, but with great phenomenological merit. However, such investigations have provided for a steady improvement of the numerical and algebraic techniques used, and thus we believe that, in conjunction with the unprecedented computer power presently available, this situation will change. First calculations of meson masses using QCD input only have already appeared \cite{Williams:2015cvx} and proof that this level of sophistication can be reached for baryons exists as well \cite{Sanchis-Alepuz:2015qra}. It is the purpose of this article to make public the numerical algorithms developed by our group  in order to enable a wider community to use and further develop them. We make no claim, however, that these same techniques are employed (or not) by other groups or whether these are the best possible ones.

The paper tackles three different classes of problems. In Sec.~\ref{sec:quark} we explain the techniques to solve the quark DSE for real and complex momenta, as needed in the calculation of bound-states. Some general guidance on the solution of DSEs for vertices is also given in Sec.~\ref{sec:quark}, although the discussion there is less detailed as they are best studied on a case-by-case basis. In Sec.~\ref{sec:mesons} the basic ideas of solving BSEs to calculate the hadron spectrum from QCD are introduced by means of the meson BSE. The BSE equation describing baryons as three-quark objects is conceptually identical to the meson BSE but computationally much more demanding; the necessary modifications to make such a calculation manageable are described in Sec.~\ref{sec:baryons}. The last important topic for which DSEs and BSEs have been used is the study of the internal structure of hadrons by calculating their form factors. The techniques here are less developed than those for the hadron spectrum; their current status is presented in Sec.~\ref{sec:formfactors}. A certain degree of repetition has been introduced on purpose, particularly in the Secs.~\ref{sec:mesons} and \ref{sec:baryons}, to make them reasonably self-contained and to enable the reader to skip to the most relevant section. We assume knowledge of the basic numerical techniques, such as e.g. quadrature and interpolation, although some formulas are quoted in the Appendices or by reference. Furthermore, although BSEs are solved as eigenvalue problems, we only sparingly discuss the relevant techniques in the text since they are part of the standard toolbox of the computational practitioner.

Finally, we wish to mention that baryons can and are treated in a simplified fashion as two-body objects, by means of introducing intermediate (unphysical) bound states called diquarks. We do not discuss the techniques used in this case here and refer instead to the relevant literature \cite{Hellstern:1997pg,Bloch:1999ke,Bloch:1999rm,Oettel:2001kd,Oettel:2000ig,Cloet:2008re,Nicmorus:2010sd,Eichmann:2011aa,Segovia:2014aza}.

%% file: 2.quarks.tex
%
\section{Quark Dyson-Schwinger equation}\label{sec:quark}
One of the fundamental ingredients to the calculation of bound-states is the constituent itself that is being bound. In this article we are concerned with bound-states of quarks and anti-quarks and hence we must provide the non-perturbative propagator for the quark. In our approach this is provided by solution of its Dyson-Schwinger equation which encodes both the perturbative and non-perturbative corrections stemming from gluonic self-interactions as well as hadronic unquenching effects.

\subsection{Setting up the equation}
\subsubsection{One-body kinematics}
The quark-propagator depends on a single momentum $p^\mu$ and can be decomposed into the tensor product of a spin-momentum part and a colour diagonal part
\begin{align}
S^{(\lambda)}_{AB}(p) = S^{(\lambda)}_{\alpha\beta}(p)\otimes\delta_{rs}\;.
\end{align}
Here, $AB$ are collective indices for spin ($\alpha\beta$) and colour ($rs$) while $\lambda$ is a flavour index that we henceforth omit in this section for brevity. 
There are two distinct Dirac structures contained within $S^{(\lambda)}_{\alpha\beta}$, namely $\hat{S}_A=-i\,\slashed{p}$ and $\hat{S}_B=\mathds{1}$, parametrized by the propagator dressing functions $\sigma_A(p^2)$, $\sigma_B(p^2)$ which are combinations of the scalar functions $A(p^2)$ and $B(p^2)$:
\begin{align}\label{eqn:quarkpropagator}
\begin{array}{rl}
S(p)  &= -i\,\slashed{p}\,\sigma_A(p^2) +     \mathds{1}\sigma_B(p^2)\\
&=   \;\;   \hat{S}_A \sigma_A(p^2) + \hat{S}_B \sigma_B(p^2)
\end{array}
\;,\;\;
\mathrm{with}\;\;\sigma_{X=A,B}(p^2) = \frac{X(p^2)}{p^2A^2(p^2) + B^2(p^2)}\;.%
\end{align}
The quark wave-function is given by $Z_f(p^2) = 1/A(p^2)$ while the quark mass-function is $M(p^2)=B(p^2)/A(p^2)$. There is an implicit dependence on the cut-off $\Lambda$ \emph{viz.} renormalization scale $\mu$.

Since the framework is covariant we are free to choose any convenient frame in which to work. The simplest choice is for the quark momentum $p^\mu$ to be aligned with that of the $\hat{e}_4$ Euclidean direction.

\subsubsection{Quark self-energy}

\begin{figure}[!b]
\begin{center}
\includegraphics[scale=0.3]{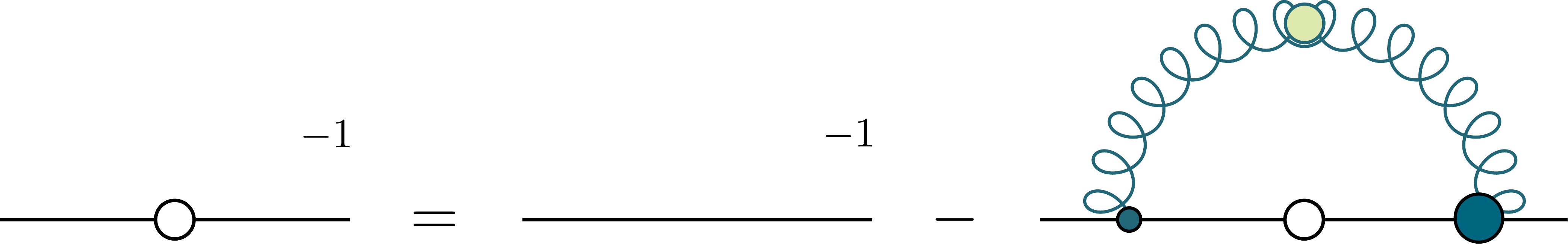}
\caption{The Dyson-Schwinger equation for the full (inverse) quark propagator. All internal propagators are dressed, with lines representing quarks and springs denoting gluons. The large blue circle is the dressed quark-gluon vertex, whilst its smaller counterpart is a bare vertex.}\label{fig:dse_quark}
\end{center}
\end{figure}

The quark-propagator is obtained by solving its Dyson-Schwinger equation, see Fig.~\ref{fig:dse_quark}. It features the bare (inverse) quark-propagator
\begin{align}
S_0^{-1}(p) = \mathrm{i} \slashed{p} + \mathds{1} Z_m m_R\;,
\end{align} 
which introduces the renormalised quark mass $m_R$ from the QCD action, and a self-energy correction $\Sigma(p^2,\mu^2)$ that encodes the dynamical quantum corrections. This self-energy contains both dressed and bare quark-gluon vertices, together with a gluon-propagator and the quark-propagator we wish to obtain. Explicitly
\begin{align}\label{eqn:dse_quark}
	S_{AB}^{-1}  (p)
	=
	Z_2S^{-1}_{AB,0}(p)
	- \Sigma_{AB}(p^2,\mu^2) \;,\quad
	\Sigma_{AB}(p^2,\mu^2)
	=
	Z_{1f}\int_k 
	\left(-ig_s\mathtt{t}^a \gamma^\mu\right)
	S(k_1)
	\left(-ig_s\mathtt{t}^b \Gamma_\mathrm{qg}^\nu(k_1,p)\right)
	D^{ab}_{\mu\nu}(k_2)\;,
\end{align}
where $k_1=k+\eta p$ and $k_2 =-k+\bar{\eta} p$ are the quark and gluon momenta, with $\eta+\bar{\eta} = 1$ expressing freedom in the momentum partitioning, and we  write the shorthand $\int_k=d^4k/\left(2\pi\right)^4$ to represent the integration measure. The quark-gluon vertices come equipped with a colour factor $\mathtt{t^a}$ which we have factored out; explicitly performing the traces (and noting that the gluon is colour diagonal) yields a global factor $C_F=(N_C^2-1)/2N_C$ in the self-energy term. The renormalisation constants $Z_2,\;Z_m$ are obtained subtractively by evaluating the quark self-energy at a given scale $\mu$ and imposing boundary conditions, typically $A(\mu^2)=1$ and $B(\mu^2)/A(\mu^2)=M(\mu^2) = m_R$. The renormalisation constant for the quark-gluon vertex is constrained by the Slavnov-Taylor identities in the miniMOM scheme~\cite{vonSmekal:2009ae}, $Z_{1f}=Z_2/\tilde{Z}_3$, where $\tilde{Z}_{3}$ renormalises the ghost propagator; in model calculations this can be taken to be equal to unity and neglected.

To solve this equation we need the explicit form of the quark-propagator Eq.~\eqref{eqn:quarkpropagator} plus two further ingredients: the (colour-reduced) quark-gluon vertex $\Gamma^\mu_\mathrm{qg}(p_1,p_2)$ and the gluon propagator $D^{ab}_{\mu\nu}(p)$. In Landau gauge, the gluon propagator has the form
\begin{align}\label{eqn:transverseprojector}
D_{\mu\nu}^{ab} = \delta^{ab} D_{\mu\nu}(p)\;,\;\;\;
D_{\mu\nu}(p) =  T_{\mu\nu}(p) D_Z(p^2)\;,\;\;\;
T_{\mu\nu}(p) = \left( \delta_{\mu\nu}-\frac{p_\mu p_\nu}{p^2}\right)\;,\;\;\;
D_Z(p^2) = \frac{Z(p^2)}{p^2}\;,
\end{align}
with $T_{\mu\nu}(p)$ the transverse projector and $Z(p^2)$ the gluon dressing function. The spin-momentum part of the quark-gluon vertex is more complicated. While its tree-level or bare form is just $\gamma^\mu$, the non-perturbative decomposition spans a linear combination of the twelve elements
\begin{align}\label{eqn:vertexanytwelve}
\gamma^\mu                          \;,\;\; p^\mu                          \;,\;\; k_1^\mu             \;,\;\;
\gamma^\mu\;\slashed{p}             \;,\;\; p^\mu\;\slashed{p}             \;,\;\; k_1^\mu\;\slashed{p}\;,\;\;
\gamma^\mu\;\slashed{k}_1             \;,\;\; p^\mu\;\slashed{k}_1             \;,\;\; k_1^\mu\;\slashed{k}_1\;,\;\;
\gamma^\mu\;\slashed{p}\;\slashed{k}_1\;,\;\; p^\mu\;\slashed{p}\;\slashed{k}_1\;,\;\; k_1^\mu\;\slashed{p}\;\slashed{k}_1\;,
\end{align}
where $k_1^\mu$ and $p^\mu$ are the incoming and outgoing quark momenta, respectively. As a result of the transverse projector inherent to Landau gauge the number of linearly independent components reduces to eight, thus we expand the vertex as $\Gamma^\mu_\mathrm{qg} = \sum_{i=1}^8 c_i(p^2,k_1^2,p\cdot k_1) \tau^\mu_i(k_1,p)$, with $c_i$ the non-perturbative and momentum dependent dressing functions and $\tau^\mu_i(k_1,p)$ an appropriate choice of basis functions, see Sec.~\ref{sec:qgvertex}.

\subsubsection{Scalar Projections}
The quark Dyson-Schwinger equation is presently a matrix equation in Dirac space. By applying a suitable projection operator, we can reduce it to a coupled set of non-linear integral equations for the scalar functions $A(p^2)$ and $B(p^2)$, viz. we multiply by
\begin{align}
P_A(p) = -i \frac{\slashed{p}}{4p^2}\;,\qquad
P_B(p) =    \frac{\mathds{1}}{4}\;,
\end{align}
and take the trace to yield
\begin{align}
A(p^2) &= Z_2\phantom{Z_M m_R}        - \mathrm{Tr} \left[ P_A(p) \Sigma(p^2,\mu^2)\right]\;, \\
B(p^2) &= Z_2 Z_M m_R - \mathrm{Tr} \left[ P_B(p) \Sigma(p^2,\mu^2)\right]\;.
\end{align}

If we split the quark propagator into its $A$ and $B$ components as in Eq.~\eqref{eqn:quarkpropagator}, we can write the trace of the spin-momentum parts of the integration kernels as
\begin{align}
K^i_{XY}(p,k) = \mathrm{Tr}\left[P_X(p) \gamma^\mu \hat{S}_Y(k_1) \tau^\nu_i(k_1,p) \right] T_{\mu\nu}(k_2)\;,
\end{align}
for $X,Y\in\left\{A,B\right\}$. Then the equations for $A$ and $B$ are
\begin{align}
    A(p^2) = Z_2\phantom{Z_M m_R}   + g_s^2 Z_{1f} C_F \int_k \frac{Z(k_2^2)}{k_2^2}\sum_{i=1}^8\left[\sigma_A(k_1^2) K^i_{AA}(p,k)+\sigma_B(k_1^2) K^i_{AB}(p,k)\right]
    c_i(k_1^2,p^2,k_1\cdot p) \;,\label{eqn:Aprojection}\\
    B(p^2) =  Z_2 Z_M m_R + g_s^2 Z_{1f} C_F \int_k \frac{Z(k_2^2)}{k_2^2}\sum_{i=1}^8\left[\sigma_A(k_1^2) K^i_{BA}(p,k)+\sigma_B(k_1^2) K^i_{BB}(p,k)\right]
    c_i(k_1^2,p^2,k_1\cdot p) \;,\label{eqn:Bprojection}
\end{align}
These traces can be performed either by hand, using FORM~\cite{Vermaseren:2000nd} or with the aid of FORMTracer~\cite{Cyrol:2016zqb}, FeynCalc~\cite{Mertig:1990an,Shtabovenko:2016sxi} and Mathematica.
Note that upon writing the integration measure in terms of hyperspherical coordinates
\begin{align}
\int_k\equiv \int\frac{d^4k}{\left(2\pi\right)^4} = 
\int_0^\infty dk^2 \frac{k^2}{2}
\int_0^\pi    d\psi\sin^2\psi 
\int_0^\pi    d\theta\sin\theta 
\int_0^{2\pi} d\phi\;.
\end{align}
the two angles $\theta$ and $\phi$ may both be trivially integrated.

\subsection{Solving the equation}
We are solving a non-linear integral equation, which requires provision of initial values for the functions of interest $A(p^2)$ and $B(p^2)$, discretized on a momentum grid $p_i^2$ distributed logarithmically in $\left[\Lambda^2_\mathrm{IR},\Lambda^2_\mathrm{UV}\right]$. For the initial guess one can for example employ
\begin{align}
A(p^2)  = \frac{1 +a}{1+p^2/\Lambda^2}\;,\;\;\;
B(p^2)  =  \frac{b}{1+p^2/\Lambda^2}\;,
\end{align}
with $a=1.2$, $b=0.8$~GeV and $\Lambda=1$~GeV being suitable choices.~\footnote{It is possible, however, to find multiple solutions of the quark Dyson-Schwinger equation, which is unsurprising since the equation is itself non-linear. Typically they have the characteristic of being \emph{noded}, in that there are one or more zero crossings present whereas the \emph{usual} solutions are positive semi-definite. These different solutions can be selected through the choice of the initial guess or by traversing different solution branches when bifurcation of the solutions occurs~\cite{Chang:2006bm,Williams:2006vva,Fischer:2008sp}.}
\begin{figure}[!ht]
\begin{center}
\includegraphics[scale=1]{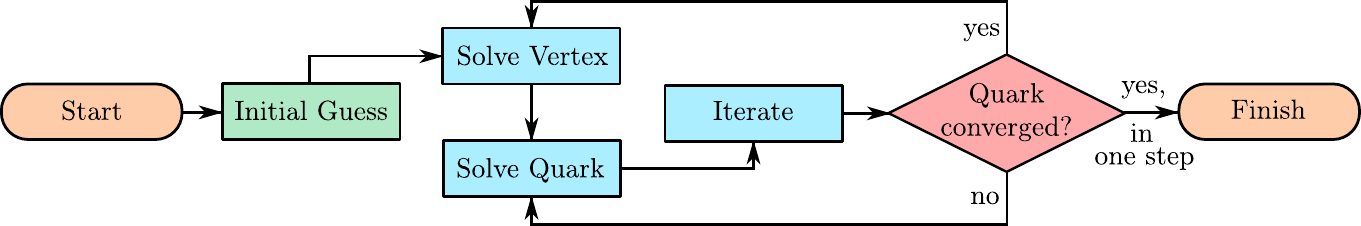}
\caption{Typical micro/macro cycle program flow for solving the quark Dyson-Schwinger equation when coupled to the quark-gluon vertex.}\label{fig:dse_quark_flowchart}
\end{center}
\end{figure}
The solution is obtained according to the flow diagram shown in Fig.~\ref{fig:dse_quark_flowchart}. That is, the quark Dyson-Schwinger equation is solved iteratively, until the two scalar functions $A$ and $B$ are converged, in the background of fixed a gluon propagator and quark-gluon vertex. This is the micro cycle; it is complemented by a macro cycle wherein the gluon propagator and quark-gluon vertex---if they are part of the coupled system---are updated to reflect their dependence on the quark propagator(s). The system is considered to have converged when further updates of the quark-gluon vertex (or gluon propagator) lead to no change in the quark propagator dressing functions (i.e. convergence is already reached in just one iteration step).

\subsubsection{Pre-calculation of the quark-gluon vertex}
Here we will keep the discussion of the quark DSE general, referring to Sec.~\ref{sec:qgvertex} for details on how the DSE for the quark-gluon vertex can be solved.
As regards its appearance in the quark DSE, the complexity is not so great, especially with respect to the memory requirements. Thus we pre-interpolate the dressing functions appearing on the right-hand side each time they are updated and cache the result.
This process is simple: for each external momentum $p^2$ required, we loop over the radial component of the integration momentum, $k^2$, together with its polar angle $\cos\psi$ and evaluate the eight dressing functions $c_i(k^2,p^2,\cos\psi)$ that describe the vertex.  In the case of rainbow-ladder this is a trivial step.

\subsubsection{Pre-calculation of angular integrals}
When the momentum configuration is chosen such that the external momentum $p$ is entirely passing through the gluon---the unshifted momentum configuration---the loop integrals can be factorized efficiently. Thus, writing $k_1= k$ and $k_2 = p-k = q$ we have for $X\in\left\{A,B\right\}$
\begin{align}\label{eqn:quarkdseangular}
    \int d^4k \frac{Z(k_2^2)}{k_2^2}\sum_{i=1}^8\sigma_Y(k_1^2) K^i_{XY}(p,k) c_i(k_1^2,p^2,k_1\cdot p)
    = \int dk^2 \frac{k^2}{2}\sigma_{Y}(k^2) \int d\psi \sin^2\psi\frac{Z(q^2)}{q^2}\sum_{i=1}^8 K^i_{XY}(p,k) c_i(k^2,p^2,q^2)\;.
\end{align}
The kernels $K^i_{XY}(p,k)$ are polynomials in $p^2$, $k^2$ and $q^2$; we eliminate $(k\cdot p)$ in favour of $q^2 = k^2 + p^2 - 2(k\cdot p)$ and collect powers in $q^2$. Suppose we also know the angular dependence of the vertex dressing functions $c_i(k^2,p^2,q^2) = \sum_j c_{ij}(k^2,p^2) P_j(\cos\psi)$ i.e. it is expanded in a basis of  orthogonal polynomials such as Chebyshev's. Then, we determine (for the whole micro cycle)
\begin{align}
\Theta_{nj}(p^2,k^2)= \int d\psi \sin^2\psi \left(q^2\right)^n Z(q^2) P_j(\cos\psi)\;,
\end{align}
with $n$ a finite set of integers, and $l\ge 0$ being the number of polynomials used in the expansion of $c_i$, whose convergence is typically rapid. This leaves
\begin{align}
\int dk^2 k^2 \sum_{Y=A,B}\sigma_Y(k^2) \sum_n\sum_{i,j} f_{XY}^{n,ij}(p^2,k^2) c_{ij}(k^2,p^2) \Theta_{nj}(p^2,k^2)\;,
\end{align}
to be solved at each step of the micro cycle, which is a single radial non-linear integral equation for $A$, $B$ alone. The $f_{XY}^{n,ij}$ are polynomials in $p^2$, $k^2$ that are determined in the course of tracing out the equations, and are related to the $K_{XY}$ appearing in Eq.~\eqref{eqn:quarkdseangular}.

For example, in the rainbow-ladder truncation where $\Gamma^\mu_\mathrm{qg}=Z_{1f}\gamma^\mu$ and $Z(p^2)=Z_\mathrm{eff}(p^2)$ is an effective interaction, the non-zero elements of $f_{XY}$ are
\begin{align}
f_{AA}^{-2} = \frac{(k^2-p^2)^2}{2p^2}  \;,\qquad\qquad
f_{AA}^{-1} =\frac{k^2+p^2}{2p^2} \;,\qquad\qquad
f_{AA}^{0}  =-\frac{1}{p^2} \;,\qquad\qquad
f_{BB}^{-1} = 3\;.
\end{align}
Then, we only need to pre-calculate the angular functions
\begin{align}
\Theta_n(p^2,k^2) = \int d\psi \sin^2\psi \left(q^2\right)^n Z_\mathrm{eff}(q^2)\;,
\end{align}
for $n=\left\{-2,-1,0\right\}$.  For the shifted momentum configuration it is not longer beneficial to pre-calculate the angular integrals since the $A$, $B$ functions that we solve for depend upon $\psi$ under the integral.

\subsection{Analytic continuation}
Everything discussed thus far can be applied to the calculation of the quark in the complex momentum plane, as needed for bound-state calculations. In this case, the choice
\begin{align}\label{eqn:quarkcomplexp}
p^\mu = \left( 0,\, 0,\, 0,\, iM + \sqrt{p^2} \right)\;,
\end{align}
results in $p^2\in\mathds{C}$ following the path of a parabola in the complex plane, with $M=0$ recovering the original real-axis parametrization. The maximal value of $M$ is dictated by the appearance of poles or singularities in the dressing functions that then require rigorous treatment. For the unshifted momentum routing ($\eta=0$, $\bar{\eta}=1$), one may either employ naive analytic continuation of the effective interaction to more cautious path deformation techniques that ensure non-analyticities are not integrated over~\cite{Alkofer:2003jj,Windisch:2016iud}. For the shifted momentum routing ($\eta=1$, $\bar{\eta}=0$), where the complex momenta is chosen to pass solely through the internal quark line, we have access to the shell and Cauchy methods~\cite{Fischer:2008sp,Krassnigg:2009gd}. Hybrid combinations of these techniques can similarly be formulated~\cite{Williams:2015cvx}. Here, we focus on the Cauchy method and note that the shell method can be thought of a nested variant.

\subsubsection{Cauchy Interpolation}
For a closed contour $\Gamma$ we can obtain $f(z_0)$ for $z_0$ interior to $\Gamma$ by the Cauchy integral formula
\begin{align}
f(z_0)& = \frac{1}{2\pi\mathrm{i}}\int_\Gamma \frac{f(z)}{z- z_0} dz\;, \\
      & = \sum_{\gamma_j} \frac{1}{2\pi\mathrm{i}}\int_{-1}^{1} \frac{f(z_j(t)) z_j^\prime(t)}{z-z_0} dt\;.
\end{align}
In the second line, we assumed that the contour is broken into a set of paths $\gamma_j$ whose points $z_j$ are parametric functions of $t\in\left[-1,1\right]$. 

This method is not stable numerically. In particular, with the contour integral is discretized at the points $z_j$, the instability manifests whenever $z_j\simeq z_0$. To counter this, we write~\cite{doi:10.1007/BF01931287,Krassnigg:2009gd}
\begin{align}\label{eqn:cauchystable}
f(z_0) =\left.\int_\Gamma \frac{f(z)}{z-z_0}dz\;\; \middle/ \;\; \int_\Gamma \frac{1}{z-z_0} dz \right.\;,
\end{align}
where the denominator evaluates to $2\pi\mathrm{i}$. In discretized form we instead write
\begin{align}\label{eqn:cauchystable_discrete}
f(z_0) =  \left.\sum_j w^\prime_j(z_0) f_j\;\; \middle/\;\; \sum_j w^\prime_j(z_0)\right.\;,\qquad w^\prime_j(z_0) = w_j / (z_0-z_j)\;.
\end{align}
where the $w_j$ are appropriate integration weights and $z_j$ the abscissa. Numerical errors when $z_0\simeq z_j$ cancel in the ratio. This form looks like barycentric interpolation discussed in~\ref{sec:barycentric}.

\subsubsection{Grid}
We can obtain the dressing functions of the quark propagator, $A$ and $B$, in the lower half-plane by complex conjugation of their values in the upper half plane. Thus it is sufficient to define only the upper half of the Cauchy contour, which is described parametrically by two paths:  the vertical part $\gamma_1$; and the parabolic part $\gamma_2$. The parametric variables $t_i\in \left[-1,1\right]$ are discretized e.g.\ as the roots of the Legendre polynomials of degree $n_i$.

\begin{figure}[!h]
	\begin{center}
		\includegraphics[scale=0.3]{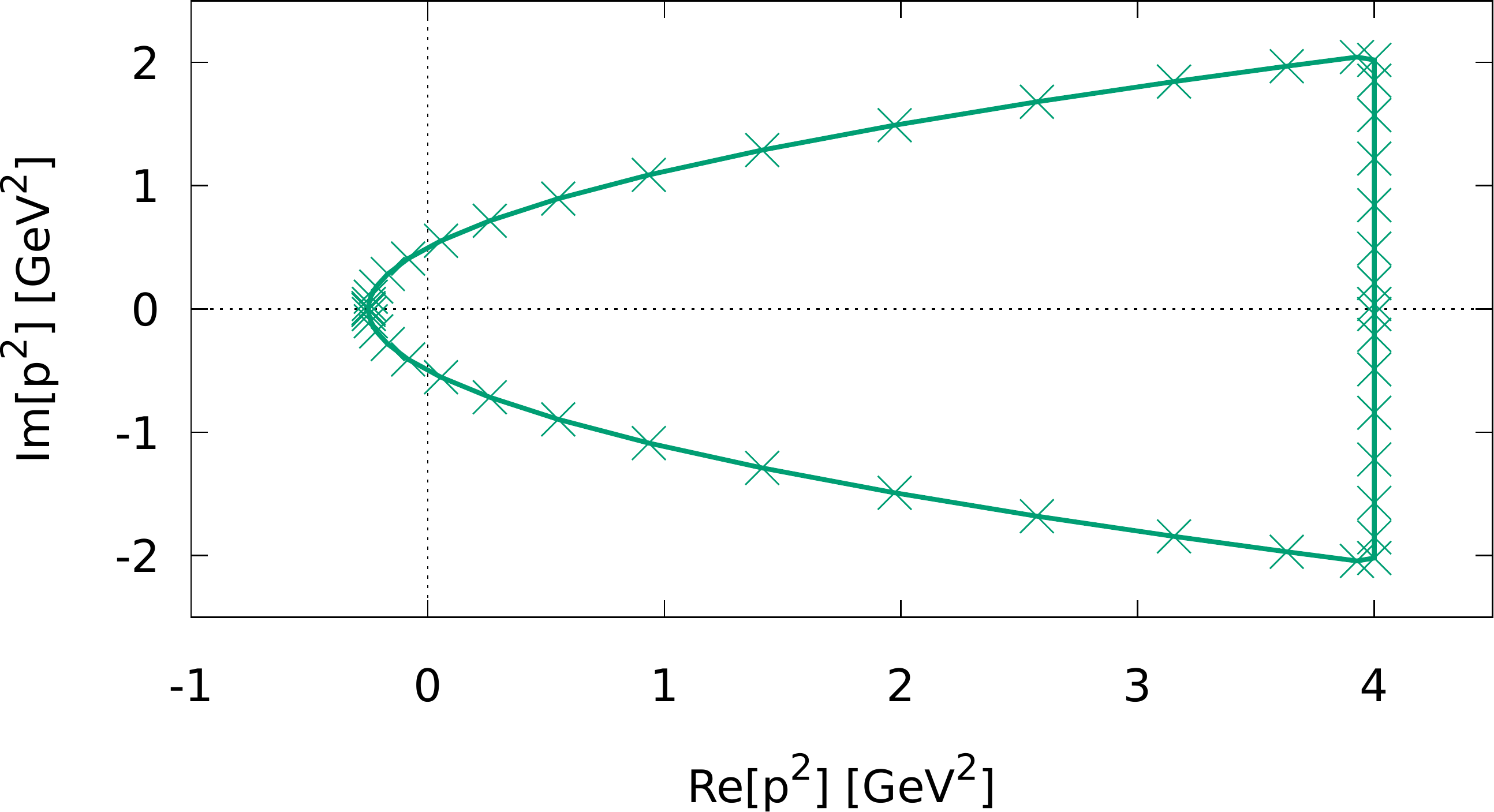}
		\caption{An example of the Cauchy integration used, for a bound-state mass $M=1$~GeV and UV cutoff $2$~GeV. The path is anticlockwise.}\label{fig:cauchycontour}
	\end{center}
\end{figure}

Defining a useful constant $c_1 = M\sqrt{4\Lambda^2+M^2}/4$, the vertical path and its derivative are given by
\begin{align}
z_{\gamma_1}(t) = \Lambda^2 + \mathrm{i}(1+t) c_1 \;,\qquad
z^\prime_{\gamma_1}(t) =  \mathrm{i} c_1\;.
\end{align}
Similarly, with the aid of $c_2 = \sinh^{-1}\left(\sqrt{4 \Lambda^2 + M^2}/2\right)/2$, the parabolic path is described with
\begin{align}
z_{\gamma_2}(t) &= \sinh^2\left(c_2 (1-t)\right) - M^2/4 + \ii M\sinh(c_2 (1-t)) \;,\nonumber \\ 
z^\prime_{\gamma_2}(t) &=- c_2 \sinh\left( 2 c_2 (1-t) \right) - \ii c_2 M \cosh\left(c_2(1-t)\right)\;.
\end{align}
We show the an example of the point distribution of this path, reflected in the plane, in Fig.~\ref{fig:cauchycontour}.

\subsection{Quark-Gluon Vertex}\label{sec:qgvertex}
The quark-gluon vertex can be decomposed into a colour part and a spin-momentum part carrying the Dirac and Lorentz indices
\begin{align}
\Gamma^{\mu,\mathtt{a}}_{ij}(k,P) = -ig_s \mathtt{t}^a \otimes \Gamma^{\mu}_{ij}\;,
\end{align}
where $\Gamma^{\mu}_{ij}$ is a linear combination of a subset of the elements given in Eq.~\eqref{eqn:vertexanytwelve}. Focusing on the spin-momentum part, the exact DSE for the quark-gluon vertex can be written as the sum 
\begin{align}\label{eqn:qgvertexcorrections}
\Gamma^\mu_\mathrm{qg}(p_1,p_2)
= Z_{1f} \left[\gamma^\mu 
+ \Lambda_\mathrm{qg,NA}^\mu
+ \Lambda_\mathrm{qg,AB}^\mu
+ \Lambda_\mathrm{qg,SF}^\mu\right]\;,
\end{align}
where the vertex corrections of the DSE are
\begin{align}
\Lambda^\mu_\mathrm{qg,NA}(p_1,p_2) &= g_s^2\phantom{\sum_{i=1}^3}\mathcal{C}_{\mathrm{NA}\phantom{,i}} 
\int \frac{d^k}{\left(2\pi\right)^4} \gamma^\alpha S(k_3) \Gamma^\beta_\mathrm{qg}(k_3,p_2) \Gamma_\mathrm{3g}^{\alpha'\beta'\mu}(k_1,-k_2,p_3)D_{\alpha\alpha'}(k_1) D_{\beta\beta'}(k_2)\;,\\
\Lambda^\mu_\mathrm{qg,AB}(p_1,p_2) &=g_s^2\phantom{\sum_{i=1}^3}\mathcal{C}_{\mathrm{AB}\phantom{,i}}  
\int \frac{d^k}{\left(2\pi\right)^4} \gamma^\alpha S(k_1)\Gamma_\mathrm{qg}^{\mu}(k_1,k_2)S(k_2) \Gamma^{\alpha'}_\mathrm{qg}(k_2,p_2) D_{\alpha\alpha'}(k_3)\;,\\
\Lambda^\mu_\mathrm{qg,SF}(p_1,p_2) & = g_s^2\sum_{i=1}^3\mathcal{C}_{\mathrm{SF},i} 
\int \frac{d^k}{\left(2\pi\right)^4} \gamma^\alpha S(k_2) \Gamma_i^{\alpha'\mu}(k_2,p_2;k_1,p_3)D_{\alpha\alpha'}(k_1)\;.
\end{align}
We have performed the colour traces, yielding the factors
\begin{align}
\mathcal{C}_\mathrm{NA} = \frac{N_C}{2}\;,\qquad
\mathcal{C}_\mathrm{AB} = -\frac{1}{2N_C}\;,\qquad
\mathcal{C}_{\mathrm{SF},1} = C_F\;,\qquad
\mathcal{C}_{\mathrm{SF},2} = -\frac{1}{2N_C}\;,\qquad
\mathcal{C}_{\mathrm{SF},3} = \frac{1}{4}\;.
\end{align}
To accomplish this we have assumed that the two-quark--two-gluon vertex can be written in colour reduced form as
\begin{align}
\Gamma^{\mu\nu,ab} 
= \mathscr{C}^{ab}_1\Gamma^{\mu\nu}_1
+ \mathscr{C}^{ab}_2\Gamma^{\mu\nu}_2
+ \mathscr{C}^{ab}_3\Gamma^{\mu\nu}_3\;,
\end{align}
with $\mathscr{C}^{ab}_1 = T^a\cdot T^b$, $\mathscr{C}^{ab}_2 = T^b\cdot T^a$, $\mathscr{C}^{ab}_3 = \delta^{ab}\mathds{1}$, where $T$ are the Gell-Mann matrices. The $\Gamma^{\mu\nu}_i$ share a common basis of $72$ Dirac-Lorentz covariants, parametrized by distinct scalar dressing functions.
The three-gluon vertex is defined with all momenta incoming. Its tree-level structure is sufficient for most applications
\begin{align}
\Gamma^{\mu\nu\rho}_{\mathrm{3g},0}(p_1,p_2,p_3) = Z_1 \left[
\delta^{\alpha\beta}\left(p_1-p_2\right)^\gamma
+\delta^{\beta\gamma}\left(p_2-p_3\right)^\alpha
+\delta^{\gamma\alpha}\left(p_3-p_1\right)^\beta
\right]\;.
\end{align}
Further details of solving the DSE for the three-gluon vertex can be found e.g.\ in Refs.~\cite{Aguilar:2013vaa,Blum:2014gna,Eichmann:2014xya,Williams:2015cvx}.

\subsubsection{Kinematics and Interpolation}\label{sec:phasespace}
\begin{figure}[!h]
	\centering\includegraphics[width=0.31\textwidth]{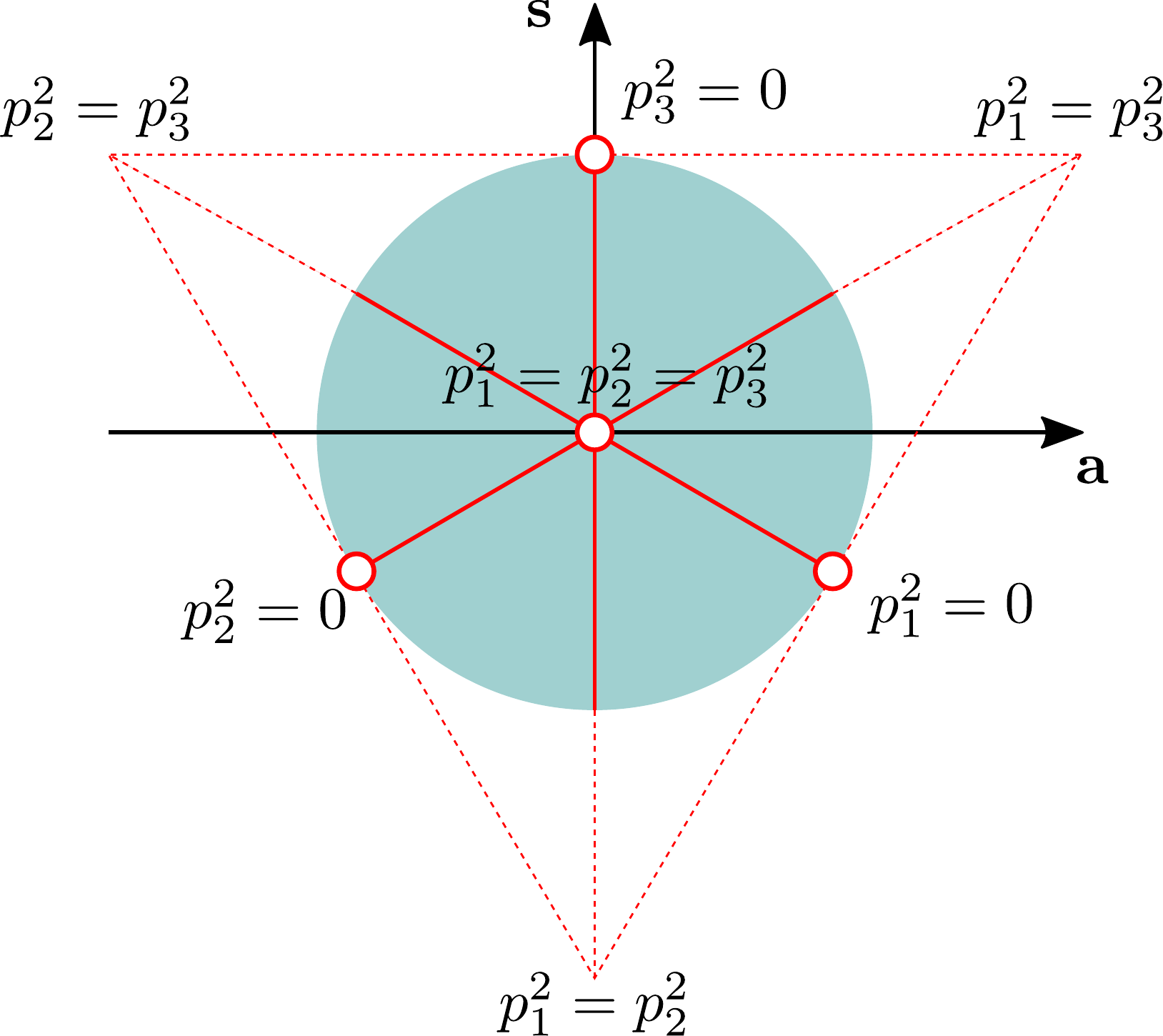}
	\caption{The $S_3$ permutation group variables arranged into a Mandelstam plane, for $s_0$ constant.}
	\label{fig:mandelstamplane}
\end{figure}
The kinematics of the quark-gluon vertex can be described by three scalar invariants corresponding to the square of the incoming and outgoing quark momenta $p_1$, $p_2$ respectively, and the incoming gluon momentum $p_3$. Due to momentum conservation, any two four-momenta are sufficient and one frequent choice is to write the relative quark momentum $k=(p_1+p_2)/2$ and the total momentum $P=p_3 = p_2-p_1$. The vertex itself may depend on any three of the scalar invariants $\left\{p_1^2,p_2^2,p_3^2,p_1\cdot p_2, p_1\cdot p_3, p_2\cdot p_3 \right\}$ or linear combinations thereof. For three-point functions 
it is convenient to arrange the three four-momenta $p_1$, $p_2$ and $p_3$ in accordance to the $S_3$ permutation group. The benefits of this have been discussed at length for the three-gluon vertex~\cite{Eichmann:2014xya} where Bose-symmetry makes its relevance obvious, but it remains useful also for the quark-gluon vertex. We write
\begin{align}\label{eqn:s0asvariables}
s_0 = \frac{p_1^2+p_2^2+p_3^2}{6}\;, \qquad
a   = \frac{\sqrt{3}\left(p_2^2-p_1^2\right)}{6s_0}\;, \qquad
s   =  \frac{p_1^2+p_2^2-2p_3^2}{6s_0}\;,
\end{align}
where the doublet $(a,s)$ forms the inside of a circle, see Fig.~\ref{fig:mandelstamplane}, whilst the singlet $s_0$ carries the overall momentum scale. In calculations, we replace this doublet by the more natural polar coordinates $(r,\psi)$, with $r\in(0,1]$ a radius and $\psi\in[0,2\pi)$ the polar angle. These variables are discretized on a grid---one that is suitable for interpolation---and the momenta $p_1$, $p_2$ and $p_3$ reconstructed. Hence
\begin{align}
\Gamma^\mu_\mathrm{qg}(p_1,p_2,p_3) = \sum_{i=1}^8 c_i \tau^\mu_i(p_1,p_2,p_3)\;,
\end{align}
with $c_i = c_i(s_0,r,\psi)$. To be concrete, $s_0$ is arranged logarithmically from $[-1,1]$ to $[\Lambda^2_\mathrm{IR},\Lambda^2_\mathrm{UV}]$ with typical scales being $10^{-6}$~GeV${}^2$ and $10^6$~GeV${}^2$. The radial variable is discretized on a grid $r^2\in[0,1]$ (thus distributing more points towards the more interesting region close to $1$). For the angular points we work in $[0,2\pi)$ with a periodic rule, offset from zero to avoid purely numerical kinematic singularities.
We use a combination of barycentric interpolation with different rules, such as Berrut and Legendre (see~\ref{sec:numerics}).

\subsubsection{Basis}
An especially useful choice is afforded by
\begin{align}\label{eqn:transversebasis}
T^{\mu\nu}(P) \Gamma_\mathrm{qg}^\nu(k,P)
&=h_1\, \gamma^\mu_T
+ h_2\, k^\mu_T \slashed{k}
+ h_3\, i k^\mu_T+ h_4\, \left(k\cdot P\right) \frac{i}{2}\left[\gamma^\mu_T,\slashed{k}\right]
+ h_5\, \frac{i}{2}\left[\gamma^\mu,\slashed{P}\right]
\nonumber\\
&+ h_6\, \frac{1}{6}\Big\{\left[ \gamma^\mu,\slashed{k}\right]\slashed{P} + \left[ \slashed{k},\slashed{P}\right]\gamma^\mu + \left[ \slashed{P},\gamma^\mu\right]\slashed{k}\Big\}+ h_7\, t^{\mu\nu}_{Pk}\left(k\cdot P\right) \gamma^\nu
+ h_8\, t^{\mu\nu}_{Pk}\frac{i}{2}\left[\gamma^\nu,\slashed{k} \right]\,.
\end{align}
Here, the incoming gluon momentum is $P^\mu$, and $k^\mu$ is the relative quark momentum.
Quantities with a subscript $T$ are contracted with the transverse projector $T^{\mu\nu}(P)$,
see \eqref{eqn:transverseprojector} and
$t^{\mu\nu}_{Pk}=\left(k\cdot P\right)\delta^{\mu\nu}-k^\mu P^\nu$.
This basis is free of kinematic singularities and transforms correctly under charge-conjugation.

\subsubsection{Numerics}\label{sec:truncation}
The calculation of the quark-gluon vertex presents a sizable numerical challenge as compared to that of propagators. Our end goal is to extract a coupled system of equations for the scalar functions that parametrize our vertex, \emph{viz.}
\begin{align}
h_i = \mathrm{Tr} \left[ P_i^\mu \Gamma^\mu \right]\;,
\end{align}
where $\Gamma^\mu$ is the DSE for the quark-gluon vertex, see Eq.~\eqref{eqn:qgvertexcorrections}, and $\mathrm{Tr}$ denotes the Dirac trace, the colour trace having already been applied. The projectors $P_i^\mu$ are defined such that $\mathrm{Tr}\left[P_i^\mu\tau_j^\mu\right]=\delta_{ij}$, and can be constructed from  linear combinations of the basis elements $\tau_j$, see Eq.~\eqref{eqn:transversebasis}. It is more efficient to determine auxiliary scalar coefficients $d_i$ by tracing with the basis elements themselves
\begin{align}
d_i = \mathrm{Tr} \left[ \tau_i^\mu \Gamma^\mu \right]\;.
\end{align}
The coefficients $h_i$ of the basis $\tau_i$ can then be reconstructed through $h_i = R_{ij} d_j$, with $R_{ij}$ the inverse of the Gram matrix $\mathrm{Tr}\left[T_i^\mu T_j^\mu\right]$. In practice this is achieved by solving the linear system of equations using LU decomposition. Any numerical issues due to an unfortunate choice of grid (i.e. where momenta become degenerate) can be determined directly from the properties of $R_{ij}$. 

It is important to choose the external momentum grid to be complementary to the basis---hence the success found in exploiting the $S_3$ permutation group variables---as this can minimize the number of numerically difficult points that are encountered. Related to this, the integrand itself must be thoroughly explored. In typical Dyson--Schwinger calculations, it suffices employ open quadrature rules such Gauss-Legendre or Gauss-Chebyshev, perhaps split into intervals \emph{e.g.} $\left(-1,0\right)\cap\left(0,+1\right)$. In the case of three-point functions, it can be that there are (integrable) end-point singularities, especially in the angle associated with the $\sin^2\theta$ part of the $d^4k$ measure in hyper-spherical coordinates; these can be dealt with through the use of the tanh-sinh rule. All of these aspects carefully chosen together, in addition to a precisely controlled interpolation of the back-coupled vertex coefficients under the integral, enables a fast and efficient calculation of this object.

\subsubsection{Analytic continuation}
The analytic continuation of the quark-gluon vertex to complex momenta is achieved in combination with the Cauchy integration method for the quark propagator. Similar to the situation there, we arrange the momentum flow inside the loop integrals of the vertex corrections such that the complex momentum associated with the total bound-state momentum $P^\mu$ such that is passes entirely through internal quark lines. 

To alleviate the problem of interpolating the quark-gluon vertex dressing functions themselves for complex momentum, we calculate them for fixed values of $M$ that appear in the $p^\mu$ of Eq.~\eqref{eqn:quarkcomplexp}. Thus we have
\begin{align}
h_i=h_i(\overline{s_0},\overline{a},\overline{s};M)\;,
\end{align}
for the dressing function, where now $\overline{s_0}$, $\overline{a}$ and $\overline{s}$ are defined as Eq.~\eqref{eqn:s0asvariables} with the four-momenta $p_i^\mu$ replaced by $\mathrm{Re}[p_i^\mu]$. The quark and vertex can be solved simultaneously for the largest desired value of $M$, thus extending the quark into the complex plane as far as possible. This is then used as input---interpolated via the Cauchy method described above---for the calculation of the quark-gluon vertex as needed in solution of the BSE.

%% file: 3.mesons.tex
%
\section{Meson Bethe-Salpeter equation}\label{sec:mesons}
The Bethe-Salpeter equation for a quark-antiquark system contains a single two-body interaction kernel that binds the quark constituents into a meson. There are innumerable  studies of the meson Bethe-Salpeter equation (see e.g. \cite{Blank:2010bp,Roberts:2011cf,Blank:2011qk,Qin:2011xq,Fischer:2014cfa,Fischer:2014xha,Hilger:2017jti} and references therein), which approach the numerical solution in several different ways. In this article we review the techniques that we employ that have been adapted to non-trivial interaction kernels beyond the rainbow-ladder approximation.

A meson in the Bethe-Salpeter formalism is described by a tensorial object $\Gamma_{AB\mathcal{I}}(p_1,p_2)$ known as the Bethe-Salpeter amplitude. The collective indices $\left\{AB\right\}$ represent the discrete spin, flavour and colour indices of the valence quarks and $\mathcal{I}=\mu_1\ldots\mu_J$ denotes the $J$ possible Lorentz indices that carry the total angular momentum of the state. The amplitude can be further decomposed into a tensor product of spin-momentum $\Psi$, flavour $F$ and colour parts
\begin{align}
\Gamma_{AB\mathcal{I}}\left(p_1,p_2\right)
=
\left( \Psi_{\alpha\beta\mathcal{I}}(p_1,p_2)\otimes F_{ab}\right)
\otimes
\frac{\delta_{rs}}{\sqrt{3}}\;,
\end{align}
where the colour part $\delta_{rs}$ fixes the bound state to be a colour singlet.

\subsection{Setting up the equation}\label{subsec:meson_settingup}

\subsubsection{Two-body kinematics}\label{subsec:meson_kinematics}
The meson Bethe-Salpeter amplitude depends on two independent momenta which could be taken as that of the two quarks, $p_1 = p+\eta P$ and $p_2 = p + (\eta-1)P$. Here $\eta\in\left[0,1\right]$ determines what fraction of the momentum $P$ passes through the quark-legs. In calculations, it is convenient to take linear combinations of these and thus form the total meson momentum $P$ and the relative quark momentum $p$. We define
\begin{align}\label{eq:mesonkinematics}
p = (1-\eta)p_1 + \eta p_2\;,\qquad\qquad P = p_1 - p_2\;.
\end{align}
We will set $\eta=1/2$ from here on.
With the total momentum fixed by the on-shell condition $P^2=-M^2$ (see below), the amplitude depends on just the relative momentum. As we will later see, it will be convenient to introduce scalar combinations of the momenta. In this case the only possibilities are 
\begin{align}
p^2\;,\;\;\; z \equiv \widehat{p}\cdot \widehat{P}\;.
\end{align}
From the two momenta $p$ and $P$, it is also convenient to introduce their transverse and orthonormal projections
$\widehat{P}$ and $\widehat{p_T}$, where $T$ denotes transverse projection with respect to the total momentum $P$ and the hat indicates normalization. In the rest frame of the meson one can choose them to be
\begin{align}
\widehat{P} = \left(0,0,0,1\right)\;,\qquad\qquad
\widehat{p_T} = \left(0,0,1,0\right)\;.
\end{align}
It will be similarly convenient to introduce transversely projected analogs of the Dirac gamma matrices and the (Euclidean space) metric tensor as this simplifies the construction of a tensor basis for the decomposition of the Bethe-Salpeter amplitude.

\subsubsection{Two-body kernel}\label{subsec:meson_kernel}
For a meson Bethe-Salpeter amplitude $\Gamma_{AB\mathcal{I}}$, the homogeneous Bethe-Salpeter equation is explicitly given by
\begin{align}
\Gamma_{AB\mathcal{I}}\left(p,P\right) = \int_k
K_{AA^{\prime},BB^{\prime}}(k_1,k_2;k)
S_{A^{\prime} A^{\prime\prime}}(k_1)
\Gamma_{A^{\prime\prime} B^{\prime\prime}\mathcal{I}}(k,P)
S_{B^{\prime\prime} B^{\prime}}(k_2)\;,
\end{align}
with the internal quark momenta being $k_1=k+P/2$ and $k_2=k-P/2$.
In general, the kernel contains spin-momentum ($K$), flavour ($k^F$) and colour ($k^C$) parts 
\begin{align}
K_{AA^\prime, B B^\prime}(k_1,k_2;k) =
K_{\alpha\alpha^\prime, \beta\beta^\prime}(k_1,k_2;k)
k^F_{aa^\prime bb^\prime}
k^C_{rr^\prime ss^\prime}\;.
\end{align}
The flavor wave function can be written as a sum of $d_F$ terms
\begin{align}
F_{ab} = \sum_\lambda^{d_F} F^{\lambda}_{ab}\;.
\end{align}
For example, the flavour content of the $K^+$ is $\left(u\overline{s}\right)$ and hence $d_F=1$, while for the $\pi^0$ we would have $d_F=2$ corresponding to $\left(u\overline{u}-d\overline{d}\right)/\sqrt{2}$.  The action of propagators on the meson amplitudes is then the sum 
\begin{align}
S_{A^{\prime} A^{\prime\prime}}(p_1)
\Gamma_{A^{\prime\prime} B^{\prime\prime}\mathcal{I}}(p_1,p_2)
S_{B^{\prime\prime} B^{\prime}}(p_2)
=
 \sum_{\lambda}^{d_F}F^{\lambda}_{ab}
S^{\left(\lambda_1\right)}_{\alpha^{\prime} \alpha^{\prime\prime}}(p_1)
\Psi_{\alpha^{\prime\prime} \beta^{\prime\prime}\mathcal{I}}(p_1,p_2)
S^{\left(\lambda_2\right)}_{\beta^{\prime\prime}\beta^{\prime} }(p_2)
\;.
\end{align}
For our example of the Kaon, the only flavour contribution yields $S^{\left(\lambda_1\right)}$ and $S^{\left(\lambda_2\right)}$ corresponding to the u- and s-quarks, respectively.

Since both the colour and flavour parts of the Bethe-Salpeter amplitude are fixed, the BSE reduces to an equation for its spin-momentum part. The final BSE is then
\begin{align}\label{eqn:finalmesonbse}
\Psi_{\alpha \beta\mathcal{I}}(p,P)
&=
\mathcal{C}\sum_{\lambda}^{d_F}\mathcal{F}^{\lambda}\int_k
K_{\alpha\alpha^\prime, \beta\beta^\prime}(k_1,k_2;k)
S^{\left(\lambda_1\right)}_{\alpha^{\prime} \alpha^{\prime\prime}}(k_1)
\Psi_{\alpha^{\prime\prime} \beta^{\prime\prime}\mathcal{I}}(k,P)
S^{\left(\lambda_2\right)}_{\beta^{\prime\prime}\beta^{\prime} }(k_2) \;,
\end{align}
where the colour trace provides a global factor $\mathcal{C}$ and contraction of the flavour parts yields
\begin{align}
\mathcal{F}^{\lambda}  =F^{\dag}_{ba} k^F_{aa^\prime bb^\prime}   F_{a^\prime b^\prime}^{,\lambda}\;.
\end{align}
Moreover, one can expand the spin-momentum part of the Bethe-Salpeter amplitude in terms of a Poincar\'e covariant tensor basis with the quantum numbers of the bound state of interest
\begin{align}\label{eqn:amplitude_expanded_in_f}
 \Psi_{\alpha\beta\mathcal{I}}(p,P)\equiv \sum_i f_i\left(p^2,z\right)~\tau^i_{\mathcal{I}}\left(p,P\right)~.
\end{align}
A covariant basis for a pseudoscalar meson is given in Eq.~\eqref{eq:pion_basis}, and its general construction discussed in~\ref{sec:mesonbasis}. The coefficients $f_i$ are scalars and can only depend on scalar combinations of the momenta.
The BSE can now be written as an equation for the scalar dressing functions
\begin{align}\label{eq:meson_BSE1}
f_i\left(p^2,z\right) = \mathcal{C}\sum_{\lambda}^{d_F}\mathcal{F}^{\lambda} g_i^{\lambda}(p^2,z)
\end{align}
where
\begin{align}\label{eq:meson_BSE2}
g_i^{\lambda}(p^2,z) = \int_k 
\bigg[
\bar{\tau}^{i}_{\beta\alpha\mathcal{I}}(p,P)
K_{\alpha\alpha^\prime, \beta\beta^\prime}(k_1,k_2;k)
S^{\left(\lambda_1\right)}_{\alpha^{\prime} \alpha^{\prime\prime}}(k_1)
\tau^{i}_{\alpha^{\prime\prime}\beta^{\prime\prime}\mathcal{I}}(k,P)
S^{\left(\lambda_2\right)}_{\beta^{\prime\prime}\beta^{\prime} }(k_2)
\bigg]f_j\left(k^2,z_k\right)\;.
\end{align}
Here $z_k=\widehat{k}\cdot \widehat{P}$ and $\bar{\tau}$ denote the conjugated basis elements, which project onto each of the elements of the covariant decomposition, i.e. $\bar{\tau}^i_{\beta\alpha\mathcal{I}}\tau^j_{\alpha\beta\mathcal{I}}=\delta^{ij}$.
Note that in Eq.~\eqref{eq:meson_BSE2} there is an implicit trace over the Dirac indices (the indices are contracted and summed over).

\subsection{Solving the Bethe-Salpeter equation}\label{sec:solving_mesonBSE}
With the BSE of a meson specified by equations \eqref{eq:meson_BSE1} and \eqref{eq:meson_BSE2}, the solution strategy is as follows:
\begin{itemize}
\item Introduce a fictitious eigenvalue $\sigma$, 
\begin{align}
\sigma~ f_i\left(p^2,z\right) = \mathcal{C}\sum_{\lambda}^{d_F}\mathcal{F}^{\lambda} g_i^{\lambda}(p^2,z)~,  
\end{align}
 to turn the BSE into an eigenvalue problem.
\item Scan the possible bound-state masses $M$ by fixing the total momentum such that $P^2=-M^2$; for example, in the rest frame of the meson this is achieved by defining it to be $P=\left(0,0,0,i~M\right)$. 
\item For each of these masses, evaluate Eq.~\eqref{eq:meson_BSE2} and solve the eigenvalue problem above. If the eigenvalue is $1$ then one has recovered the original BSE and has solved the problem. The bound-state mass is the corresponding $M$ and the eigenvectors give the dressing functions $f(p^2,z)$. If the eigenvalue is different from $1$, try a different bound-state mass and refine the search using a secant method.
\end{itemize}

Some details and references on how to solve eigenvalue problems in the context of BSEs are given in~\ref{sec:eigenvalues}. In this section we focus on how we optimally calculate Eq.~\eqref{eq:meson_BSE2}.

\subsubsection{Preliminaries}\label{subsec:mesonpreliminaries}
Each dressing function depends on one radial and one angular variable. For a meson, to evaluate \eqref{eq:meson_BSE2} one does not need to employ additional interpolation (though it may be beneficial for various reasons) since they can be mapped directly to the integration variables. To do so the four-dimensional integration \eqref{eq:meson_BSE2} is performed using hyper-spherical coordinates $\{k^2,z,y,\phi\}$
\begin{align}
\int_k\equiv \int\frac{d^4k}{\left(2\pi\right)^4} = \int dk^2 \frac{k^2}{2}\int dz \sqrt{1-z^2}\int dy\int d\phi\;.
\end{align}
The integration over $\phi$ is trivial and gives just a numerical factor. For the radial variable and the angle $y$ we employ Gauss-Legendre quadrature rules. For the variable $z$ we use Gauss-Chebyshev quadrature as it minimises the number of necessary integration points. Furthermore, the quadrature points for the radial variables must be mapped from $[-1,1]$ to $[0,\infty )$. A logarithmic mapping to $[\Lambda_{IR},\Lambda_{UV}]$ with $\Lambda_{IR}\sim 10^{-3}$ and $\Lambda_{UV}\sim 10^{3}$ is sufficiently well-behaved.

\subsubsection{Pre-calculation of the wave function}\label{subsec:mesonwavefunction}
It is convenient to work instead with the Bethe-Salpeter wave function, $\Phi$, which may be obtained from the amplitude by attaching free quark propagators
\begin{align}
 \Phi_{\alpha\beta}^\lambda(p,P)= S^{\left(\lambda_1\right)}_{\alpha \alpha^{\prime}}(p_1)
\Psi_{\alpha^{\prime} \beta^{\prime}\mathcal{I}}(p,P)
S^{\left(\lambda_2\right)}_{\beta^{\prime}\beta }(p_2)\;.
\end{align}
It can be expanded in the same covariant basis as the amplitudes
\begin{align}
\Phi_{\alpha\beta}^\lambda(p,P)\equiv\sum_i \omega^{\lambda}_i (p^2,z) \tau^i_{\alpha\beta\mathcal{I}}(p,P)~.
\label{eq:meson_wave_function}
\end{align}
The coefficients $f^i$ of Eq.~\eqref{eqn:amplitude_expanded_in_f} and $\omega^i$ can be related by means of a \textit{rotation} matrix $Y$
\begin{align}
\omega^{\lambda}_i (p^2,z)= Y^{ij}_{(\lambda_1\lambda_2)}(p,P) f^j(p^2,z)~.
\end{align}
\begin{itemize}
	\item \textbf{Pre-calculation of the propagator traces}: To expand the wave function in terms of the covariant basis we need to evaluate
	\begin{align}
	Y^{ij}_{(\lambda_1\lambda_2)}(k,P) = 
	\bar{\tau}^i_{\beta\alpha\mathcal{I}}(k,P)
	S^{(\lambda_1)}_{\alpha\alpha^\prime}(k_1)
    \tau^j_{\alpha^\prime\beta^\prime\mathcal{I}}(k,P)
    S^{(\lambda_2)}_{\beta^\prime\beta}(k_2)
	\end{align}
    For the meson covariant basis the traces are not complicated and the scalar rotation matrix $Y^{ij}_{(\lambda_1\lambda_2)}(k,P)$ can be pre-calculated for the integration grid in $(k^2,\, z_k)$. This requires just a few MB of storage.
\end{itemize}

\subsubsection{Pre-calculation of the kernel}\label{subsec:mesonkernel}
Once the wave function has been expanded in the covariant basis, one is left with the evaluation of the following trace of the BSE interaction kernel
\begin{align}\label{eqn:mesonkerneltrace}
L^{ij}(p,k;P)=
\bar{\tau}_{\beta\alpha\mathcal{I}}^i(p,P)
K_{\alpha\alpha^\prime,\beta\beta^\prime}(p,k;P)
\tau_{\alpha^\prime\beta^\prime\mathcal{I}}^j(k,P)
\end{align}
The traces can be performed algebraically in advance and evaluated as required. Some degree of pre-calculation can be employed depending upon the choice of the interaction kernel $K$. For example, in the simple case of a rainbow-ladder kernel the following steps are convenient:
\begin{itemize}
\item \textbf{Separation of the traces and the momentum dependence}:
In the rainbow-ladder kernel the tensor structure is $\gamma^\mu\otimes\gamma^\nu$ together with a momentum-dependent transverse projector for the gluon exchange and a momentum-dependent coupling
\begin{align}
L^{ij}(p,k;P)&=
\left[
\bar{\tau}_{\beta\alpha\mathcal{I}}^i(p,P)
\gamma^\mu_{\alpha\alpha^\prime}
\tau_{\alpha^\prime\beta^\prime\mathcal{I}}^j(k,P)
\gamma^\nu_{\beta^\prime\beta^\prime}
\right] D_{\mu\nu}(k-p)\nonumber \\
&=L^{ij}_{\mu\nu}D_{\mu\nu}(k-p)\;.
\end{align}
	If we make use of the following transverse orthonormal momenta
	\begin{align}
	\widehat{P}^\mu = ( 0,\,0,\,0,\, 1)\;,\qquad
	\widehat{p_T}^\mu = ( 0,\,0,\,1,\, 0)\;,\qquad
	\widehat{k_T}^\mu = ( 0,\,\sin\theta,\,\cos\theta,\, 0)\;,
	\end{align}
	with $\theta$ one of the hyper-spherical angles (and its cosine $z=\cos\theta$ is used as integration variable) in \eqref{eq:meson_BSE2}, then a possible covariant basis for the Bethe-Salpeter amplitude is given by
	\begin{align}\label{eq:pion_basis}
	\tau^1 &= \gamma_5\;,\;\;\;\;
	\tau^2 = \phantom{-}\mathrm{i}\gamma_5\widehat{\slashed{k}_T}\;,\;\;\;\;
	\tau^3 = \phantom{-}\mathrm{i}\gamma_5\slashed{\widehat{P}}\;,\;\;\;\;
	\tau^4 = \gamma_5\widehat{\slashed{k}_T}\slashed{\widehat{P}}\;,\nonumber\\
	\overline{\tau}^1 &= \gamma_5\;,\;\;\;\;
	\overline{\tau}^2 = -\mathrm{i}\widehat{\slashed{p}_T}\gamma_5\;,\;\;\;\;
	\overline{\tau}^3 = -\mathrm{i}\slashed{\widehat{P}}\gamma_5\;,\;\;\;\;
	\overline{\tau}^4 = \slashed{\widehat{P}}\widehat{\slashed{p}_T}\gamma_5\;.
	\end{align}
	This basis corresponds to a pseudoscalar meson, \emph{e.g.} a pion. Then, we would find for the kernel traces
	\begin{equation}
	\begin{aligned}[c]
	L_{\mu\nu}^{11}(\theta) &= \delta_{\mu\nu}\;,\;\;\nonumber \\
	L_{\mu\nu}^{14}(\theta) &= \widehat{k_T}_\mu \widehat{P}_\nu - \widehat{k_T}_\nu \widehat{P}_\mu\;,\;\;\nonumber \\
	L_{\mu\nu}^{22}(\theta) &= \widehat{k_T}_\mu \widehat{p_T}_\nu  + \widehat{k_T}_\nu \widehat{p_T}_\mu  - \delta_{\mu\nu}\widehat{k_T}\cdot \widehat{p_T}\;,\;\;\nonumber \\
	L_{\mu\nu}^{23}(\theta) &= \widehat{p_T}_\mu \widehat{P}_\nu  + \widehat{p_T}_\nu \widehat{P}_\mu\;,\;\;\nonumber
	\end{aligned}
	\begin{aligned}[c]
	L_{\mu\nu}^{32}(\theta) &= \widehat{k_T}_\mu \widehat{P}_\nu  + \widehat{k_T}_\nu \widehat{P}_\mu\;,\;\;\nonumber \\
	L_{\mu\nu}^{33}(\theta) &= 2\widehat{P}_\mu \widehat{P}_\nu  - \delta_{\mu\nu}\;,\;\;\nonumber \\
	L_{\mu\nu}^{41}(\theta) &= \widehat{p_T}_\mu \widehat{P}_\nu  - \widehat{p_T}_\nu \widehat{P}_\mu\;,\;\;\nonumber \\
	L_{\mu\nu}^{44}(\theta) &=  - \widehat{k_T}_\mu \widehat{p_T}_\nu  - \widehat{k_T}_\nu \widehat{p_T}_\mu  - 2\widehat{P}_\mu \widehat{P}_\nu \widehat{k_T}\cdot \widehat{p_T} + \delta_{\mu\nu}\widehat{k_T}\cdot \widehat{p_T}
	\end{aligned}
	\end{equation}
	with the remaining matrix elements vanishing. The only dependence is on the angle $\theta$ owing to the simplicity of the rainbow-ladder interaction and the use of transverse orthonormal momenta for the external/internal basis.	
	\item \textbf{Factorisation of the angular dependence}:
	The trace in Eq.~\eqref{eqn:mesonkerneltrace} is not overly complicated for meson bound-states. Moreover, this is the only place where the combination $p\cdot k$ can appear and exposes the dependence of the kernel on the integration variable $y$. We define $q = k-p$ and eliminate $p\cdot k$ in favour of $q^2$. Then, collecting powers of $q^2$ allows us to factorise with respect to the angular variable: this integral we can pre-calculate.
	      More-or-less independent of the structure of the kernel we then need
	      \begin{align}
	      G^x_y(p,k;P) = \int_{-1}^{1}dy \left(q^2\right)^x h_y(q^2;\ldots)\;,
	      \end{align}
	      where $x\in\left\{-2,J+1\right\}$---for total angular momentum $J$---is the power to which $q^2$ is raised and $h_y(q^2)$ are the set of scalar functions contained within the kernel that depend on $q^2$ (and hence the angular variable $y$), such as the gluon propagator dressing or the coefficients of any internal quark-gluon vertices.
\end{itemize}

%% file: 4.baryons.tex
%
\section{Baryon Bethe-Salpeter equation}\label{sec:baryons}
\begin{figure}[!ht]
\begin{center}
\includegraphics[scale=0.15]{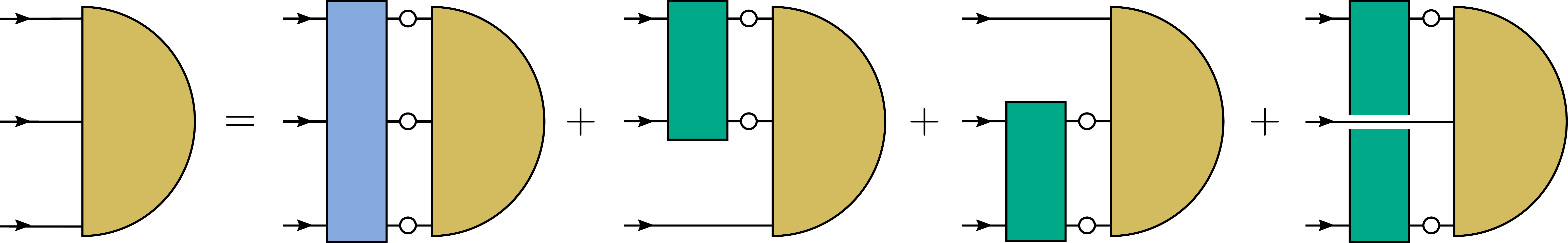}
\caption{Schematic representation of a baryon BSE. Green boxes represent two-body irreducible interaction kernels whereas the blue box represents a three-body irreducible kernel. Lines with blobs denote fully-dressed quark propagators.}\label{fig:baryon_BSE}
\end{center}
\end{figure}

The baryon and meson Bethe-Salpeter equations are similar in many respects and for this reason many concepts discussed in this section will be a repetition of those in Sec.~\ref{sec:mesons}. The BSE for a three-quark system contains two types of diagrams (see Fig.\ref{fig:baryon_BSE}), those containing a two-body interaction kernel and a spectator quark and those with a three-body interaction kernel. The effect of the three-body kernel is expected to be small and is usually omitted. Moreover, until very recently \cite{inpreparation:threebody}, it has been considered intractable in a covariant approach. When the three-body kernel is neglected, the simplified BSE is usually called the Faddeev equation: this is the only case that we consider here.
The basic procedure to solve the Faddeev equation was outlined in~\cite{Eichmann:2009qa,Eichmann:2011vu} (see also \cite{Eichmann:2009zx}). We present here the key ideas for a numerical implementation of the Faddeev equation with a \emph{generic} two-body interaction kernel, most of which are not really necessary for simplistic kernels such as rainbow-ladder.
\subsection{Setting up the equation}

\subsubsection{Three-body kinematics}\label{subsec:baryon_kinematics}
The baryon Bethe-Salpeter amplitude depends on three independent momenta. They could be taken as the quark momenta $p_1$, $p_2$ and $p_3$. As we saw in previous sections, however, it is very convenient in many cases to rewrite them in terms of the total baryon momentum $P$ and two relative momenta $p$ and $q$ since then, for a fixed total momentum, the amplitude effectively depends on two momenta only. We define\footnote{As in Sec.~\ref{subsec:meson_kinematics}, a momentum partitioning parameter $\eta$ can be introduced. We will only work with $\eta=\nicefrac{1}{3}$, so we omit it here.}
\begin{align}\label{eq:defpq}
        p &= 2\,p_3/3 - (p_1+p_2)/3\,,
               &  p_1 &=  -q -\dfrac{p}{2} + P/3\,, \nonumber\\
        q &= \left( p_2-p_1\right)/2\,,
               &  p_2 &=   q -\dfrac{p}{2} + P/3\,, \nonumber\\
        P &= p_1+p_2+p_3\,,
               &  p_3 &=   p + P/3\,\,,
\end{align}
When the spin-momentum part of the Bethe-Salpeter amplitude is expanded in a Poincar\'e covariant tensor basis (see~\ref{sec:numerics}), then the coefficients of such an expansion (the dressing functions) can only depend on scalar combinations of the momenta. One can choose them to be
\begin{align}\label{eq:momentum_scalars}
p^2\;,\qquad
q^2\;,\qquad
z_0\equiv \widehat{p_T}\cdot \widehat{q_T}\;,\qquad
z_1\equiv \widehat{p}\cdot \widehat{P}\;,\qquad
z_2\equiv \widehat{q}\cdot \widehat{P}\;,
\end{align}
with the $T$ denoting transverse projection with respect to the total momentum.

Another advantage of this choice of momenta is made manifest when used in combination with a transverse and orthonormal covariant basis. We call in this section a basis \emph{transverse orthonormal} if---in addition to being orthonormal with respect to an inner product space that we discuss later---it is further composed of unit momentum vectors, transverse with respect each other, \emph{e.g.} the momenta $\widehat{P}$, $\widehat{p_T}$ and $\widehat{q_{TT}}$, with ${}_T$ denoting transverse projection with respect to the total momentum $P$ and ${}_{TT}$ transverse projection with respect to both $P$ and $p_T$.

Moreover, since the framework is covariant, we are free to choose any reference frame. For the Faddeev equation it is convenient to work in the baryon rest frame, where $\widehat{P}$, $\widehat{p_T}$ and $\widehat{q_{TT}}$ can be taken to define three reference axes (in hyperspherical coordinates)
\begin{align}\label{eq:rest_frame_unit_transverse}
\widehat{P}=\left( 0,0,0,1 \right)\;,\qquad
\widehat{p_T}=\left( 0,0,1,0 \right)\;,\qquad
\widehat{q_{TT}}=\left( 0,1,0,0 \right)\;.
\end{align}
As we see below, in such a frame the integration momentum takes a very simple form when used in a transverse orthonormal basis.

Note, however, that such a choice of momenta is not optimally behaved under permutations of the quark labels, which plays an essential role in reducing the Faddeev equation to something manageable.

\subsubsection{Relation between the different diagrams}\label{subsec:three_diagrams_from_one}
Even without the three-body kernel, the calculation of the covariant Faddeev equation with its full momentum dependence---once we take into account the large number of covariants necessary to describe the amplitude---is very demanding. A crucial step forward was taken in Ref.~\cite{Eichmann:2011vu}, where it was shown that for a certain class of two-body kernels, only one diagram in the Faddeev equation needs to be calculated explicitly, while the remaining two can be reconstructed from the former.  The proof follows after realising that the colour and flavour parts of the Bethe-Salpeter amplitude enforce certain symmetry properties on the spin-momentum part. Specifically, a baryon is described by the three-body Bethe-Salpeter amplitude $\Gamma_{ABCD}(p,q,P)$, where $\{ABC\}$ are collective indices for spin, flavour and colour indices for the valence quarks and, similarly, we use $D$ as a collective index for the resulting baryon. Since it must describe a system of three fermions, it must be antisymmetric under the exchange of any two quark labels. When we decompose it as a tensor product of spin-momentum, flavour and colour parts we can fix the last two to be
\begin{align}\label{eq:BSE_amplitude}
 \Gamma_{ABCD}(p,q,P)=\left(\sum_\rho
\Psi^\rho_{\alpha\beta\gamma\mathcal{I}}(p,q,P) \otimes
F^\rho_{abcd}\right)\otimes \frac{\epsilon_{rst}}{\sqrt{6}}~,
\end{align}
where the colour term $\epsilon_{rst}/\sqrt{6}$ constrains the baryon to be a colour singlet
and the flavour terms $F^\rho_{abcd}$ are the quark-model $\mathrm{SU}(N_f)$-symmetric, antisymmetric or mixed irreducible
representations; the index $\rho$ is relevant only for the mixed representations where we must sum over both mixed-symmetric and mixed-antisymmetric components. From this it is clear that the product of the spin-momentum and flavour terms must be symmetric under the exchange of quark labels. For a given baryon, the symmetry of the flavour amplitudes is known and thus determines the symmetry of the spin-momentum part.
Now, in detail the Faddeev equation reads
\begin{align}\label{eq:symbolic_faddeev_original}
   \Gamma_{ABC\mathcal{I}}(p,q,P)={}&\int_k~\Bigl[~\textrm{K}_{BB',CC'}(k_2,\widetilde{k}
_3;k)~\delta_
{AA''}~\textrm{S}_{B'B''}(k_2)~\textrm{S}_{C'C''}(\widetilde{k}_3)\Gamma_{
A''B''C''\mathcal{I}}(p^{(1)},q^{(1)},P)~\Bigr]~+ \nonumber \\
&\int_k~\Bigl[~\textrm{K}_{AA',CC'}(\widetilde{k}_1,k_3,k)~\delta_
{BB''}~\textrm{S}_{A'A''}(\widetilde{k}_1)~\textrm{S}_{C'C''}(k_3)\Gamma_{
A''B''C''\mathcal{I}}(p^{(2)},q^{(2)},P)~\Bigr]~+\nonumber\\
&\int_k~\Bigl[~\textrm{K}_{AA',BB'}(k_1,\widetilde{k}_2;k)~\delta_
{CC''}~\textrm{S}_{A'A''}(k_1)~\textrm{S}_{B'B''}(\widetilde{k}_2)\Gamma_{
A''B''C''\mathcal{I}}(p^{(3)},q^{(3)},P)~\Bigr]~.
  \end{align}
The kernel also contains spin-momentum, flavour and colour parts
\begin{align}
\textrm{K}_{AA',BB'}(p_1,p_2;p)=K_{\alpha\alpha',\,\beta\beta'}(p_1,p_2;p)k^F_{
aa'bb'}k^C_{rr',\,ss'}~.
\end{align}
Moreover, we write the flavour wave functions as a sum of several terms
\begin{align}
 F^\rho_{abcd}=\sum_\lambda^{d_F} F^{\rho,\lambda}_{abcd}\;,
\end{align}
where $d_F$ is the number of such terms (for example, the antisymmetric
component of the $\Xi^0$ is $(uss-sus)/\sqrt{2}$ and therefore $d_F=2$). The
action of the propagators on the Faddeev amplitudes in
(\ref{eq:symbolic_faddeev_original}) can therefore be written as a sum of $d_F$
terms, for instance
\begin{align}
\textrm{S}_{AA'}(p_1)&\textrm{S}_{BB'}(p_2)\Gamma_{A'B'C\mathcal{I}}(p,q,P)=\sum_\rho \sum_i^{d_F}S^{(\lambda_1)}_{\alpha\alpha'}(p_1)
S^{(\lambda_2)}_{\beta\beta'}(p_2)
\Psi^\rho_{\alpha'\beta'\gamma\mathcal{I}}(p,q,P)F^{\rho,i}_{abc}\;,
\end{align}
where, for example, for the first term in the antisymmetric component of the
$\Xi^0$ given above, $S^{(\lambda_1)}$ and $S^{(\lambda_2)}$ would be the
propagators for $u$- and $s$-quarks, respectively.
The quark propagators depend on the internal quark momenta
$k_i=p_i-k$ and $\tilde{k}_i=p_i+k$, with $k$ the exchanged momentum. The
internal relative momenta, for each of the three terms in the Faddeev equation, are
\begin{align}\label{internal-relative-momenta}
  p^{(1)} &= p+k\;,
& p^{(2)} &= p-k\;,
& p^{(3)} &= p  \;,\nonumber\\
  q^{(1)} &= q-k/2\;,
& q^{(2)} &= q-k/2\;,
& q^{(3)} &= q+k\,\,.
\end{align}
Now, from Eq.~\eqref{internal-relative-momenta} it is clear that for the third term in Eq.~\eqref{eq:symbolic_faddeev_original} the internal and external relative momenta $p$ coincide. This makes this term particularly easy to evaluate, since for the integration we will need to interpolate the Faddeev amplitude in the $q$-direction only. Therefore, our goal is to relate the first two terms in Eq.~\eqref{eq:symbolic_faddeev_original} to the third one.  It is obvious that the first and second terms can be obtained from the third by permutations of the quark labels $\{123\}\rightarrow\{231\}$ and $\{123\}\rightarrow\{132\}$, respectively. In particular, the kinematics of the first two terms can be expressed as in the third one, but evaluated at shifted relative momenta
\begin{align}
\textrm{1st diagram:}\qquad& \left\{p,q\right\}
~~\to ~~
\left\{p'=-q-\frac{p}{2}\;,\;\; q'=-\frac{q}{2}+\frac{3p}{4}\right\}\;, \label{eq:shifted_momenta_1}\\
\textrm{2nd diagram:}\qquad& \left\{p,q\right\}
~~\to ~~
\left\{p''=q-\frac{p}{2}\;,\;\; q''=-\frac{q}{2}-\frac{3p}{4}\right\}\;.\label{eq:shifted_momenta_2}\\
\end{align}
Putting in all the elements defined above and permuting the indices in
Eq.~\eqref{eq:symbolic_faddeev_original} as indicated above, and after renaming dummy indices
conveniently, Eq.~\eqref{eq:symbolic_faddeev_original} reduces now to an equation for the spin-momentum parts of the Faddeev amplitude
\begin{align}\label{eq:faddeev_transformed}
\Psi^\rho_{\alpha\beta\gamma\mathcal{I}}(p,q,P)=&\,C~\mathcal{F}_1^{\rho\rho',
\lambda}\int_k~\Bigl[~K_{\beta\beta',\gamma\gamma'}(k'_1,\widetilde{k}'
_2;k)~\delta_{\alpha\alpha''}~S^{(\lambda_2)}_{\beta'\beta''}(k'_1)~S^{
(\lambda_3)}_{\gamma'\gamma''}(\widetilde{k}'_2)\;\Psi^{\rho'}_{
\beta''\gamma''\alpha'' \mathcal{I}}(p'^{(3)},q'^{(3)},P)~\Bigr] \nonumber \\
+&\,C~\mathcal{F}_2^{\rho\rho',\lambda}\int_k~\Bigl[~K_{\alpha\alpha',\gamma\gamma'
}(k''_1,\widetilde{k}''
_2;k)~\delta_{\beta\beta''}~S^{(\lambda_1)}_{\gamma'\gamma''}(\widetilde{k}
''_1)~S^{(\lambda_3)}_{\alpha'\alpha''}(k''_2)\;\Psi^{\rho'}_{
\gamma''\alpha''\beta'' \mathcal{I}}(p''^{(3)},q''^{(3)},P)~\Bigr]\nonumber\\
+&\,C~\mathcal{F}_3^{\rho\rho',\lambda}\int_k~\Bigl[~K_{\alpha\alpha',\beta\beta'}
(k_1,\widetilde{k}
_2;k)~\delta_{\gamma\gamma''}~S^{(\lambda_1)}_{\alpha'\alpha''}(k_1)~S^{
(\lambda_2)}_{\beta'\beta''}(\widetilde{k}_2)\;\Psi^{\rho'}_{
\alpha''\beta''\gamma''\mathcal{I}}(p^{(3)},q^{(3)},P)~\Bigr]\;,
\end{align}
where the contraction of the color parts of the kernel and those of the Faddeev
amplitudes (before the permutation of indices) gives a global factor $C$ and,
similarly, the contraction of the corresponding flavour parts leads to the flavour
matrices
\begin{align}\label{eq:flavor_matrices}
\mathcal{F}_1^{\rho\rho',\lambda}=F^{\dag~\rho}_{bac}k^F_{bb'cc'}F^{\rho',
\lambda}_{b'c'a}~,\qquad
\mathcal{F}_2^{\rho\rho',\lambda}=F^{\dag~\rho}_{bac}k^F_{aa'cc'}F^{\rho',
\lambda}_{c'a'b}~,\qquad
\mathcal{F}_3^{\rho\rho',\lambda}=F^{\dag~\rho}_{bac}k^F_{aa'bb'}F^{\rho',
\lambda}_{a'b'c}~.
\end{align}
If we denote the result of the integral in the third line of Eq.~\eqref{eq:faddeev_transformed} as
$\left[\Psi^{(3)}_{\lambda_1\lambda_2}\right]^{\rho}_{\alpha\beta\gamma
\mathcal{I}}(p,q,P)$, it is now clear that if the kernel is such that
$K_{\alpha\alpha',\,\beta\beta'}(k_1,k_2,k)=K_{\beta\beta',\,\alpha\alpha'}(k_1,k_2,
k)$ then we have
\begin{align}\label{eq:faddeev_transformed_compact}
\Psi^\rho_{\alpha\beta\gamma
	\mathcal{I}}(p,q,P)=\;C~\mathcal{F}_1^{\rho\rho',\lambda}\left[\Psi^{(3)}_{\lambda_2\lambda_3}(p',q',P)\right]
^{\rho'}_{\beta\gamma\alpha
	\mathcal{I}}~
+\;C~\mathcal{F}_2^{\rho\rho',\lambda}\left[
\Psi^{(3)}_{\lambda_1\lambda_3}(p'',q'',P)\right]^{\rho'}_{\gamma\alpha\beta
	\mathcal{I}}~
+\;C~\mathcal{F}_3^{\rho\rho',\lambda}\left[
\Psi^{(3)}_{\lambda_1\lambda_2}(p,q,P)\right]^{\rho'}_{\alpha\beta\gamma\mathcal{I}}~.
\end{align}
Therefore, the task at hand has been reduced to the calculation of only one of the
diagrams in the Faddeev equation for all the necessary combinations of pairs of
quarks. As in Sec.~\ref{sec:mesons}, we expand the spin-momentum part in a covariant tensor basis $\tau_{\alpha\beta\gamma\mathcal{I}}(p,q,P)$ which encodes the spin of the baryon \cite{Eichmann:2009qa,SanchisAlepuz:2011jn}. One can then rewrite Eq.~\eqref{eq:faddeev_transformed_compact} in terms of the scalar
coefficients $f(p^2,q^2,z_0,z_1,z_2)$ of the expansion of the Faddeev amplitudes in such a covariant
basis. These coefficients (the dressing functions) depend on scalar combinations of the momenta only. We finally obtain an equation for the scalar dressing functions
\begin{align}\label{eq:Faddeev_coeff}
 f^{\rho}_i(p^2,q^2,z_0,z_1,z_2)
 &=C\mathcal{F}_1^{\rho\rho';\lambda}~H_1^{ij}~g^{\rho',\lambda}_{j}(p'^2,q'^2,z'_0,z'_1,z'_2)\nonumber\\
 &+C\mathcal{F}_2^{\rho\rho';\lambda}~H_2^{ij}~g^{\rho',\lambda}_{j}(p''^2,q''^2,   z''_0,z''_1,z''_2)\nonumber\\
 &+C\mathcal{F}_3^{\rho\rho';\lambda}~\phantom{H_3^{ij}}~g^{\rho',\lambda}_{i}(p^2,q^2,z_0,z_1,z_2)~,
\end{align}
with
\begin{align}\label{eq:Faddeev_coeff3}
g^{\rho,\lambda}_i(p^2,q^2,z_0,z_1,z_2)=\int_k~\Bigl[~\bar{\tau}^
{i}_{\beta\alpha\mathcal{I}\gamma}(p,q,P)K_{\alpha\alpha',\beta\beta'}(p,q,
k)~&\delta_{\gamma\gamma''}S^{(\lambda_1)}_{\alpha'\alpha''}(k_1)~S^{(\lambda_2)
}_{\beta'\beta''}(k_2)~\tau^{j}_{\alpha''\beta''\gamma''\mathcal{I}}(p^{(3)},q^{
(3)},P)~\Bigr] \nonumber\\&
\times f^{\rho}_j(p_{(3)}^2,q_{(3)}^2,z^{(3)}_0,z^{(3)}_1,z^{(3)}_2)~,
\end{align}
and
\begin{align}\label{eq:rotation_matrices}
H_1^{ij}=\left[\bar{\tau}^i_{\beta\alpha\mathcal{I}\gamma}(p,q,P)\tau^j_{
\beta\gamma\alpha\mathcal{I}}(p',q',P)\right]~,\qquad
H_2^{ij}=\left[\bar{\tau}^i_{\beta\alpha\mathcal{I}\gamma}(p,q,P)\tau^j_{
\gamma\alpha\beta\mathcal{I}}(p'',q'',P)\right]~,
\end{align}
where $\bar{\tau}^i$ denotes the \textit{conjugation} of (or projection onto) the $i$-th basis element, conjugation appropriately defined such that the inner product yields $\bar{\tau}^i_{\beta\alpha\mathcal{I}\gamma}\tau^j_{\alpha\beta\gamma\mathcal{I}}=\delta^{ij}$ ~\cite{Eichmann:2009qa,SanchisAlepuz:2011jn}.

\subsection{Solving the Faddeev equation}\label{subsec:Faddeev_tricks}
As explained at the beginning of Sec.~\ref{sec:solving_mesonBSE}, BSEs are solved as eigenvalue problems by introducing a fictitious eigenvalue and scanning the space of bound-state masses $M$ (e.g., by fixing the total momentum to be $P=(0,0,0,\mathrm{i}~M)$ in the rest frame of the baryon) until finding a value for which the eigenvalue is one. In this section we describe the most important steps necessary to implement the numerical evaluation of Eq.~\eqref{eq:Faddeev_coeff}. Note that most of what is explained here is not strictly necessary for simple kernels such as in the rainbow-ladder approximation, in which case the techniques described in Sec.\ref{sec:solving_mesonBSE} to solve a meson BSE together with the relation among the three different diagrams explained above, suffice to solve the Faddeev equation with modest computer resources. It is for more complicated interaction kernels that the techniques described below are useful, if not unavoidable.

\subsubsection{Preliminaries}\label{subsubsec:generalities}
As already stated above, we will use the variables in Eq.~\eqref{eq:momentum_scalars} to describe the momentum dependence of the scalar dressing functions; that is, each dressing function depends on two radial variables and three angular variables. Note that in \cite{Eichmann:2011vu} it is argued that after rewriting the radial variables in terms of a permutation-group singlet and a doublet (a member of one), the dependency of the dressings on the latter is weak, thus facilitating the interpolation. We don't consider this possibility here.

One needs to interpolate over these variables. For the radial variables the interpolation methods of choice are cubic spline or barycentric interpolation (see \ref{sec:barycentric}). Whilst the latter is somewhat faster, it generates slight \textit{oscillations} in the high-momentum tail of the amplitudes, which can be problematic if those amplitudes are later used to calculate \emph{e.g.} form factors. For the angular variables, it is customary to expand them in Chebyshev polynomials (see \ref{sec:chebyshev}). This expansion then defines a natural interpolation in these variables. Another possibility is to use barycentric interpolation for the angular variables as well, which can be defined in such a way that it is equivalent to the Chebyshev expansion.

Finally, a four-dimensional integration over the variables in \eqref{eq:q3_integration_variable} must be performed. Experience shows that Gauss-Legendre quadrature rules are optimal for the radial variable as well as the angles $y$ and $\phi$. For the angle $z$, Gauss-Chebyshev quadrature is more efficient requiring fewer integration points. Furthermore, the quadrature points for the radial variables must be mapped from $[-1,1]$ to $[0,\infty )$. A logarithmic mapping to $[\Lambda_\mathrm{IR},\Lambda_\mathrm{UV}]$ with $\Lambda_\mathrm{IR}\sim 10^{-3}$ and $\Lambda_\mathrm{UV}\sim 10^{3}$ is sufficiently well-behaved.

\subsubsection{Pre-calculation of the wave function}
Instead of solving Eq.~\eqref{eq:Faddeev_coeff3} directly, it is convenient to attach the quark propagators to the Faddeev amplitude outside of the integral. Since the resulting wave function just represents the free propagation of quarks, it will have the same quantum numbers as the Faddeev amplitude and can therefore be expanded in the same covariant basis, \emph{e.g.}
\begin{align}
\Phi^{\rho;\lambda_1\lambda_2}_{\alpha\beta\gamma\mathcal{I}}(p,q,P)\equiv S^{(\lambda_1)}_{\alpha\alpha'}(p_1)
S^{(\lambda_2)}_{\beta\beta'}(p_2)
\Psi^\rho_{\alpha'\beta'\gamma\mathcal{I}}(p,q,P)\equiv\sum_i \omega^{\rho;\lambda_1\lambda_2}_i (p^2,q^2,z_0,z_1,z_2) \tau^i_{\alpha\beta\gamma\mathcal{I}}(p,q,P)~,
\label{eq:baryon_wave_function}
\end{align}
and the wave function carries an additional index (a pair of them) indicating the flavour of the quarks that have been attached to the amplitude.
One then solves the following integral instead
\begin{align}\label{eq:Faddeev_coeff3_wave}
g^{\rho,\lambda}_i(p^2,q^2,z_0,z_1,z_2)=\!\int_k~\Bigl[~\bar{\tau}^
{i}_{\beta\alpha\mathcal{I}\gamma}(p,q,P)K_{\alpha\alpha',\,\beta\beta'}(p,q,
k)~&\delta_{\gamma\gamma'}\,\tau^{j}_{\alpha'\beta'\gamma'\mathcal{I}}(p^{(3)},q^{
(3)},P)~\Bigr]\times \omega^{\rho;\lambda_1\lambda_2}_j (p_{(3)}^2,q_{(3)}^2,z^{(3)}_0,z^{(3)}_1,z^{(3)}_2)~.
\end{align}
To arrive here we proceed as follows:
\begin{itemize}
\item \textbf{Pre-calculation of the propagator traces}: To expand the wave function in terms of the covariant basis one needs to evaluate traces of the following type
\begin{align}
\bar{\tau}^
{i}_{\beta\alpha\mathcal{I}\gamma}(p,q,P)S^{(\lambda_1)}_{\alpha\alpha'}(p_1)
S^{(\lambda_2)}_{\beta\beta'}(p_2)~\tau^{j}_{\alpha'\beta'\gamma\mathcal{I}}(p,q,P)
.\label{eq:propagator_dirac_trace}
\end{align}
Traces involving baryon covariant bases are computationally expensive, so it is convenient to pre-calculate and store them. However, Eq.~\eqref{eq:propagator_dirac_trace} is both momentum and quark-flavour dependent through the quark dressing functions and hence too big to be prestored. We prestore instead
\begin{align}
D_{ij}^{\mu\nu}=&
 ~\bar{\tau}^
{i}_{\beta\alpha\mathcal{I}\gamma}(p,q,P)\gamma^{\mu}_{\alpha\alpha'}\gamma^{\nu}_{\beta\beta'}~\tau^{j}_{\alpha'\beta'\gamma\mathcal{I}}(p,q,P)
~.\label{eq:propagator_dirac_trace_MI}
\end{align}
with $\mu,~\nu$ running from $0$ to $4$ and $\gamma^0$ denoting the unit matrix. If, as discussed in Sec.~\ref{subsec:baryon_kinematics}, the basis depends on the momenta via $\widehat{P}$, $\widehat{p_T}$ and $\widehat{p_{TT}}$ only, then in the reference frame Eq.~\eqref{eq:rest_frame_unit_transverse}, $D_{ij}^{\mu\nu}$ is effectively momentum independent so it can be calculated in advance, requiring just a few MB of memory. Then, Eq~\eqref{eq:propagator_dirac_trace} is reconstructed from $D_{ij}^{\mu\nu}$ in the next step.

\item \textbf{Evaluation of $\omega$ at the integration points}: For each fixed total momentum $P$, the propagator traces \eqref{eq:propagator_dirac_trace} need to be evaluated anew since the quark dressing functions depend on $P$ through the quark momentum. This can be done outside of the main integration step. After evaluating the dressings $\sigma_s(p_i)$ and $\sigma_v(p_i)$, we can form a 5-dimensional vector $d(p_i)=(\sigma_s,\sigma_v p_i^1,\sigma_v p_i^2,\sigma_v p_i^3,\sigma_v p_i^4)$ and the traces in Eq.~\eqref{eq:propagator_dirac_trace} can be recovered with a simple matrix-vector multiplication (omitting flavour indices)
\begin{align}
 \bar{\tau}^
{i}_{\beta\alpha\mathcal{I}\gamma}(p,q,P)S^{(\lambda_1)}_{\alpha\alpha'}(p_1)
S^{(\lambda_2)}_{\beta\beta'}(p_2)~\tau^{j}_{\alpha'\beta'\gamma\mathcal{I}}(p,q,P)
=d^\mu (p_1) d^\nu (p_2) D_{ij}^{\mu\nu}~.
\end{align}
Next, since we only need to solve the diagram \eqref{eq:Faddeev_coeff3_wave} and due to the simplified kinematics in this case (see Eqs.~\eqref{eq:shifted_momenta_1}
and \eqref{eq:shifted_momenta_2}), we can use $q^{(3)}$ as integration variable. In hyperspherical coordinates it would read
\begin{align}
q^{(3)}=\sqrt{q^2}~\lb \sin\phi \sqrt{1-z^2}\sqrt{1-y^2},~\cos\phi \sqrt{1-z^2}\sqrt{1-y^2},~\sqrt{1-z^2}~y,~z \rb~,
\label{eq:q3_integration_variable}
\end{align}
and there is a one-to-one correspondence between some of the integration variables and the momentum scalars on which the dressings of Faddeev amplitude (and of the wave function) depends. Namely
\begin{align}
\left( p^{(3)} \right)^2=p^2\;,\qquad
\left( q^{(3)} \right)^2=q^2\;,\qquad
z^{(3)}_0=y\;,\qquad
z^{(3)}_1=z_1\;,\qquad
z^{(3)}_2=z\;.
\end{align}
Therefore, we can directly evaluate the wave-function dressings $\omega$ at the integration points needed for Eq.~\eqref{eq:Faddeev_coeff3_wave}, thus avoiding interpolation under the integral. As before, only a few hundred MB are necessary to prestore the wave-function dressings.
\end{itemize}

\subsubsection{Pre-calculation of the kernel}
The next computationally expensive element in the integral of Eq.~\eqref{eq:Faddeev_coeff3_wave} is the trace over the interaction kernel
\begin{align}
\bar{\tau}^
{i}_{\beta\alpha\mathcal{I}\gamma}(p,q,P)K_{\alpha\alpha',\beta\beta'}(p,q,
k=q^{(3)}-q)~\delta_{\gamma\gamma'}~\tau^{j}_{\alpha'\beta'\gamma'\mathcal{I}}(p^{(3)},q^{(3)},P)
~.\label{eq:kernel_trace_initial}
\end{align}
As with the propagator traces, it is convenient to pre-calculate and store it. This is, however, infeasible for any minimally sophisticated interaction kernel. For example, for the kernel studied in \cite{Sanchis-Alepuz:2015qra} which consisted of a single dressed-gluon exchange, a direct pre-calculation and storage of the kernel would have required memory in the PB range. This problem can be overcome as follows:
\begin{itemize}
\item \textbf{Separation of the traces and the momentum dependence}: One of the reasons for the high memory requirements is the large dimension of a typical covariant baryon basis. For example, for a spin-$\nicefrac{1}{2}$ baryon (the lowest-dimensional possibility), the indices $i,j$ above run both from $1$ to $64$. Therefore the first step is to factorise as much as possible the momentum dependence and the tensor structure of the kernel. This is trivial (as much as unnecessary) for a rainbow-ladder kernel, where the kernel consists of a $\gamma^\mu\otimes\gamma^\nu$ tensor part, contracted with a momentum-dependent transverse projector and possibly other momentum-dependent couplings. It is not so trivial in general and probably a case-by-case strategy must be used. For example, the kernel in Ref.~\cite{Sanchis-Alepuz:2015qra} was of the type $\Gamma^\mu\otimes\gamma^\mu$, $\Gamma^\mu$ being the fully-dressed quark-gluon vertex (with full momentum dependence) with the transverse projector included. A full quark-gluon vertex can be expanded in 12 (momentum-dependent) tensor structures. However, in \cite{Sanchis-Alepuz:2015qra} it was expanded in a 64-dimensional (corresponding to the dimension of an arbitrary object with two Dirac indices and a Lorentz index) momentum-independent basis. The use of such a redundant and seemingly counter-intuitive basis was the crucial step to make the calculation feasible. To see this, let us assume that we can factorise our kernel as
\begin{align}
 K_{\alpha\alpha',\,\beta\beta'}(p,q,k)=T^\kappa_{\alpha\alpha',\,\beta\beta';\delta}~\lambda^{\kappa;\delta}(p,q,k)~,
\end{align}
with $T^\kappa$ being a set of momentum-independent tensors, $\delta$ a superindex subsuming possible discrete indices (such as the indices $\mu$, $\nu$ for the example of a rainbow-ladder kernel) and $\lambda^\kappa$ a set of momentum-dependent dressing functions. The trace \eqref{eq:kernel_trace_initial} then factorises as
\begin{align}
\left[~\bar{\tau}^
{i}_{\beta\alpha\mathcal{I}\gamma}(p,q,P)T^\kappa_{\alpha\alpha',\beta\beta';\delta}~\delta_{\gamma\gamma'}~\tau^{j}_{\alpha'\beta'\gamma'\mathcal{I}}(p_{(3)},q_{
(3)},P)~\right]~\lambda^{\kappa;\delta}(p,q,k=q^{(3)}-q)~,\label{eq:kernel_trace_split}
\end{align}
and the only momentum dependence of the trace comes now from the basis elements. However, using the frame \eqref{eq:rest_frame_unit_transverse} for the \textit{external} momenta $p$, $q$, the momentum $q^{(3)}$ as integration variable as defined in Eq.~\eqref{eq:q3_integration_variable} and assuming that we work with a transverse orthonormal basis, the only momentum dependence left in the trace comes from
\begin{align}
\widehat{q^{(3)}_{TT}}=\left( \sin\phi,~\cos\phi,~ 0 ,~0\right)\;,
\end{align}
that is, only a dependence on the angle $\phi$ remains. This can easily be pre-calculated and stored.

It is preferable as well---for each value of the total momentum $P$---to pre-calculate and store the momentum-dependent parts $\lambda^{\kappa;\delta}$ outside the integration since, in general, their evaluation can easily become the bottleneck of the calculation. The memory requirements to do that are high but realistic, ranging from a few GB to $1$--$2$~TB depending on the number of integration points and the complexity of the kernel. In this way, in the integration, only the \textit{reconstruction}  of the kernel via the contraction of the $\kappa$ and $\delta$ indices in Eq.~\eqref{eq:kernel_trace_split} needs to be done.
\end{itemize}

\subsubsection{Reconstruction of the remaining diagrams}
This is the most costly part for a rainbow-ladder kernel, but its cost becomes marginal for more complicated choices. As can be inferred from Eq.~\eqref{eq:Faddeev_coeff}, the necessary steps are, for each value of the momentum scalars $\{p^2,q^2,z_0,z_1,z_2\}$, to interpolate the dressing functions for the shifted momenta \eqref{eq:shifted_momenta_1} and \eqref{eq:shifted_momenta_2} as well as calculate the rotation matrices \eqref{eq:rotation_matrices}. This last step can be slightly alleviated by providing analytic expressions for them (obtained, for example, with a symbolic-computing software such as Mathematica).

\subsubsection{Choice of basis and its impact on the numerics}
It is clear that any tensor basis with the correct quantum numbers is equally valid to describe a baryon (or a meson, or a vertex) and, in principle, to be used to decompose the BSE into a set of equations for scalar dressing functions. This, however, does not necessarily imply that any basis is equally well suited for the numerical implementation of those equations. For example, a possible orthonormal basis for a spin-$\nicefrac{3}{2}$ baryon is given by the 128 elements
\begin{align}
\sqrt{\frac{3}{2}}
\bracket{\gamma_5 \Gamma^i\bracket{p,q}\Lambda\bracket{\pm}\gamma_5 C}_{\alpha\beta}
\bracket{\phantom{\gamma_5}\Gamma^{j\phantom{,\,\mu}}\bracket{p,q}\Lambda\bracket{+}p^\mu~\mathcal{P}^{\mu\nu}}_{\gamma\delta}~,\nonumber\\
\sqrt{\frac{3}{2}}
\bracket{\phantom{\gamma_5}\Gamma^i\bracket{p,q}\Lambda\bracket{\pm}\gamma_5 C}_{\alpha\beta}
\bracket{\gamma_5 \Gamma^{j\phantom{,\,\mu}}\bracket{p,q}\Lambda\bracket{+}p^\mu~\mathcal{P}^{\mu\nu}}_{\gamma\delta}~,\nonumber\\
\sqrt{2}
\bracket{\gamma_5 \Gamma^i\bracket{p,q}\Lambda\bracket{\pm}\gamma_5 C}_{\alpha\beta}
\bracket{\phantom{\gamma_5}\tilde{\Gamma}^{j,\,\mu}\bracket{p,q}\Lambda\bracket{+}\phantom{p^\mu~}\mathcal{P}^{\mu\nu}}_{\gamma\delta}~,\nonumber\\
\sqrt{2}
\bracket{\phantom{\gamma_5}\Gamma^i\bracket{p,q}\Lambda\bracket{\pm}\gamma_5 C}_{\alpha\beta}
\bracket{\gamma_5 \tilde{\Gamma}^{j,\,\mu}\bracket{p,q}\Lambda\bracket{+}\phantom{p^\mu~}\mathcal{P}^{\mu\nu}}_{\gamma\delta}~,
\label{eq:direct_ONbasis}
\end{align}
with $\mathcal{P}^{\mu\nu}$ the Rarita-Schwinger projector for spin-$\nicefrac{3}{2}$ particles~\cite{Rarita:1941mf}, $C$ the charge conjugation matrices, $\Lambda\bracket{\pm}=\left(\mathds{1}\pm\widehat{\slashed{P}}\right)/2$ and
\begin{align}
\Gamma\bracket{p,q}\equiv& \left\{ \mathds{1}, \frac{1}{2}\left[\pslash,\qslash\right],\pslash,\qslash \right\}~,\\
\tilde{\Gamma}^{\mu}\bracket{p,q}\equiv& \left\{ \Gamma^1~q^\mu-\frac{1}{2}\Gamma^2~p^\mu,~ \Gamma^2~q^\mu+\frac{1}{2}\Gamma^1~p^\mu,~\Gamma^3~q^\mu-\frac{1}{2}\Gamma^4~p^\mu,~\Gamma^4~q^\mu+\frac{1}{2}\Gamma^3~p^\mu \right\}~,\label{eq:vectorchoiceofbasis}
\end{align}
with the form of~\eqref{eq:vectorchoiceofbasis} the result of Gram-Schmidt orthonormalisation.

However, this is an extremely poorly-behaved basis for numerical calculations. One of the reasons is that all basis elements are dependent on the relative momenta, whereas a ground-state baryon is typically dominated by s-waves which, in this framework, are described by momentum-independent basis elements. These must be recovered from the basis \eqref{eq:direct_ONbasis} by means of complicated linear combinations of basis elements with an equally complicated momentum dependence of the coefficients (dressing functions). A basis which already encodes at least some of the expected physics is therefore preferable. In this respect, bases with a well-defined partial-wave content (and, as a consequence, well-behaved numerically) have been constructed for spin-$\nicefrac{1}{2}$ \cite{Eichmann:2009qa}, spin-$\nicefrac{3}{2}$ \cite{SanchisAlepuz:2011jn} and spin-$\nicefrac{5}{2}$ \cite{52_in_prep} baryons.

%% file: 5.formfactors.tex
\section{Form factors}\label{sec:formfactors}

Once the meson or baryon BSE has been solved one has access---not only to the hadron spectrum---but to its internal structure as encoded in the Bethe-Salpeter amplitude or wave function. In order to extract this information, one needs to couple the hadron to external probes. The most studied case in the literature is the coupling of hadrons to external electromagnetic fields and the calculation of the corresponding current, from which the electromagnetic form factors of hadrons can be determined. We detail the corresponding numerical techniques here (however, the same techniques apply for the extraction of other types of form factors~\cite{Eichmann:2011pv}). Note also that baryons and their form factors can be studied using a quark-diquark approximation of the three-body equation, which we don't consider here (see, for example~\cite{Hellstern:1997pg,Bloch:1999ke,Bloch:1999rm,Oettel:2001kd,Oettel:2000ig,Cloet:2008re,Nicmorus:2010sd,Eichmann:2011aa,Segovia:2014aza}).

\subsection{Normalisation}
Since BSEs are homogeneous equations of the BS amplitude, its normalisation is not fixed. Thus, before using the amplitudes in the calculation of currents their physical normalisation must be determined; then \emph{a priori} current conservation can be realised self-consistently. One can establish two different but equivalent normalisation conditions, whose derivation we only sketch here. One can start from the equation for the scattering matrix $T$, introducing an auxiliary variable $\vartheta$ which will correspond to the inverse of the eigenvalue to be determined in the solution of the BSE. Namely,
\begin{align}
T=K+\vartheta K G_0 T \qquad\to\qquad T=\bracket{1-\vartheta K G_0}^{-1}~K~.
\end{align}
Let us now take a derivative with respect to $\vartheta$ and thereafter make further use of the equation above
\begin{align}
\frac{dT}{d\vartheta}=\bracket{1-\vartheta K G_0}^{-1}~KG_0~\bracket{1-\vartheta K G_0}^{-1}~K=T G_0 T~.\label{eq:Nakanishi0}
\end{align}
We introduce the BS amplitudes by means of the expansion of $T$ around the bound-state pole
\begin{align}
T\sim\frac{\Psi\bar{\Psi}}{P^2+M^2\bracket{\vartheta}}~,
\end{align}
where we have assumed that the dependence on the parameter $\vartheta$ enters through the bound-state mass $M$ only. Expanding Eq.~\eqref{eq:Nakanishi0} we find
\begin{align}
-\frac{\Psi\bar{\Psi}}{\bracket{P^2+M^2\bracket{\vartheta}}^2}\frac{dM^2}{d\vartheta}=\frac{\Psi\bar{\Psi}G_0\Psi\bar{\Psi}}{\bracket{P^2+M^2\bracket{\vartheta}}^2}
\qquad\to\qquad
 -\frac{dM^2}{d\vartheta}=\bar{\Psi}G_0 \Psi~.
\end{align}
We can now rewrite this in terms of the BSE eigenvalue $\lambda=\vartheta^{-1}$, taking into consideration that the BSE solution corresponds to the case $\lambda=1$,
\begin{align}
\left.\frac{dM^2}{d\lambda}\right|_{\lambda=1}=\bar{\Psi}G_0 \Psi~.\label{eq:Nakanishi_normalisation}
\end{align}
This is the Nakanishi normalisation condition for BS amplitudes~\cite{Nakanishi:1965zza}. It is the simplest normalisation condition to use in practice, since it only requires  one to save the eigenvalues for different masses in the process of solving the BSE, and then the evaluation of a four- (for mesons) or eight-dimensional (for baryons) integration over the relative momenta, implicit in Eq.~\eqref{eq:Nakanishi_normalisation}. For example, in the case of mesons
\begin{align}
\left.\frac{dM^2}{d\lambda}\right|_{\lambda=1}=\int_q \bar{\Psi}_{\beta\alpha\mathcal{I}}\bracket{q,P^2=-M^2}~S_{\beta\beta'}\bracket{q+P/2}S_{\alpha'\alpha}\bracket{q-P/2}~\Psi_{\alpha'\beta'\mathcal{I}}\bracket{q,P^2=-M^2}~.
\end{align}
It can be checked that the Nakanishi~\cite{Nakanishi:1965zza} and the Leon-Cutkosky~\cite{Cutkosky:1964zz} normalisation conditions are equivalent~\cite{Fischer:2008wy,Fischer:2009jm}.

\subsection{Calculation of hadronic currents}
\begin{figure}[!ht]
\begin{center}
\hspace*{\fill}
\includegraphics[scale=0.15]{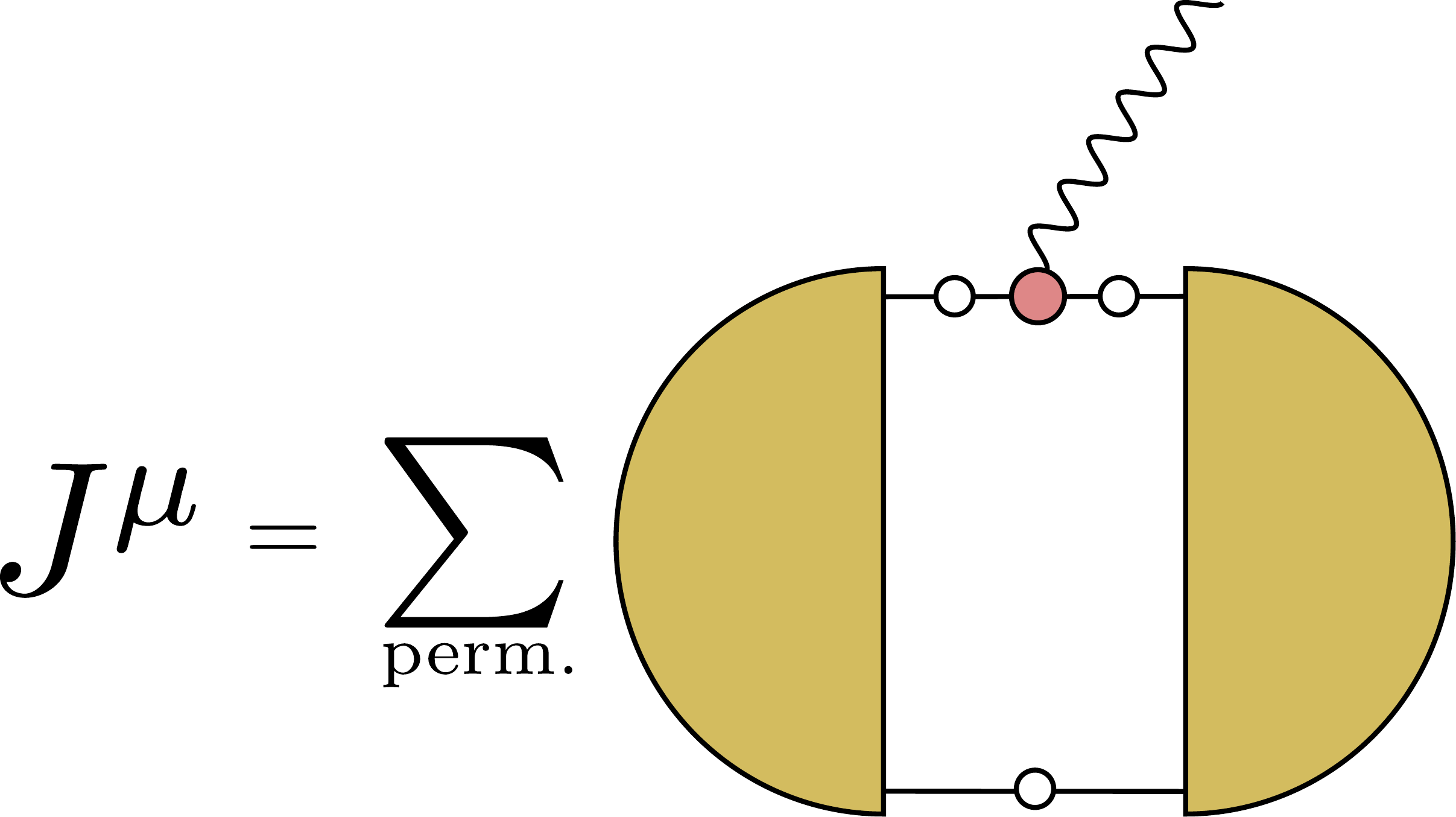}\hfill
\includegraphics[scale=0.15]{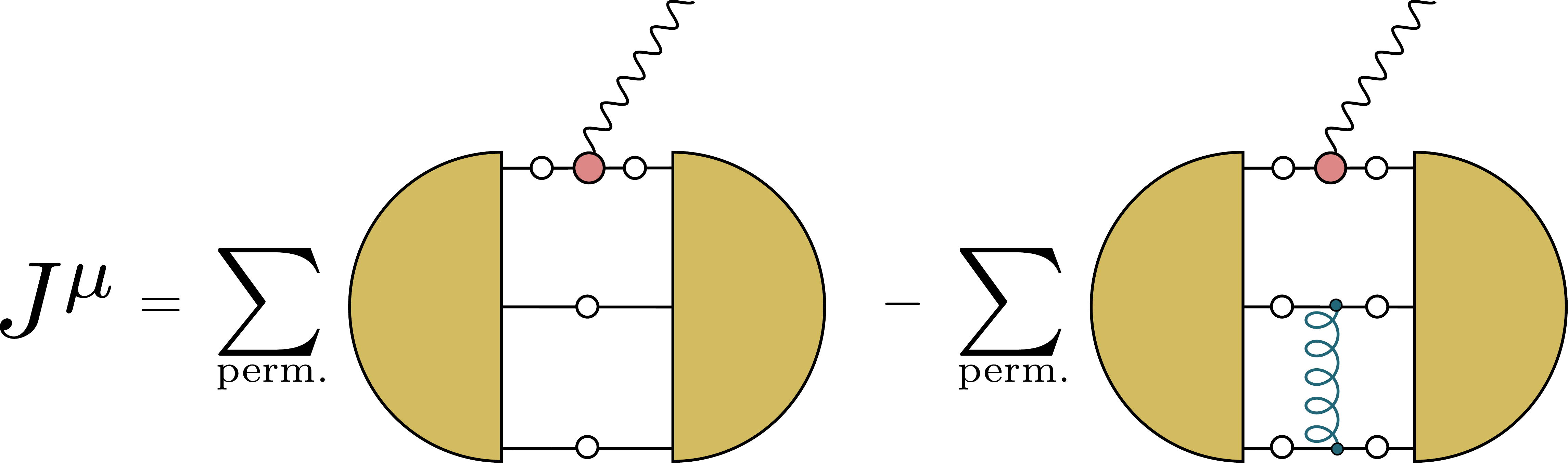}\hspace*{\fill}
\caption{Diagrams necessary for the calculation of the current in Eq.~\eqref{eq:FFeq_compact} for the rainbow-ladder truncation, for the case of mesons (left panel) and baryons (right panel). The permutations refer to the different ways of attaching the external field to the quark propagator lines.}\label{fig:FFs_diagrams}
\end{center}
\end{figure}

For a hadron, calculated using a BSE with interaction kernel $K$, the current $J^\mu$ describing its coupling to an external field is given by (see, e.g. \cite{Eichmann:2011ec})
\begin{align}\label{eq:FFeq_compact}
 J^\mu_{\mathcal{I'}\mathcal{I}}=\bar{\Psi}_\mathcal{I'} \left(G_0^\mu-iG_0 K^\mu G_0\right)\Psi_\mathcal{I}~,
\end{align}
where the \textit{gauged} Green's functions $G_0^\mu$ and $K^\mu$ are defined as follows \cite{Kvinikhidze:1998xn}. The product of quark propagators encodes the coupling of the external field to the constituent quarks
\begin{align}
G_0^\mu=S^\mu~S\cdots+S~S^\mu\cdots+\dots~,
\end{align}
via the definition of the quark-photon vertex $\Gamma^\mu$
\begin{align}
S(p_f-p_i)^\mu\equiv S(p_f) \Gamma^\mu(p_f-p_i) S(p_i)~.
\label{eq:quark_vertex}
\end{align}
The gauged kernel $K^\mu$ may contain, depending on whether one is dealing with mesons or baryons, two type of contributions. The kernel $K$ is decomposed in terms of $\ell$-particle irreducible kernels $K^{(\ell)}$ (with $\ell\ge 2$)
\begin{align}
K= \sum_\ell K^{(\ell)} \underbrace{S^{-1}\cdots S^{-1}}_{N-\ell}~,
\end{align}
with $N=2,3$ for mesons and baryons, respectively. The \textit{gauged} kernel then consists of
\begin{align}
K^\mu=\sum_\ell  [K^{(\ell)}]^\mu S^{-1}\cdots S^{-1}+K^{(\ell)} [S^{-1}]^\mu\cdots S^{-1}+\dots~,
\end{align}
with $[S^{-1}]^\mu=-S^{-1}~S^\mu~S^{-1}$. That is, it contains additional quark-photon couplings for the spectator quark lines (if any), plus the terms $[K^{(\ell)}]^\mu$ in which the external field couples to the internal lines of the interaction kernel. For a rainbow-ladder kernel, in which the only internal line corresponds to a neutral gluon, $[K^{(\ell)}]^\mu$ vanishes. To the best of our knowledge, all calculations performed so far assumed $[K^{(\ell)}]^\mu=0$ and this is the only possibility we consider here. A diagrammatic representation of Eq.~\eqref{eq:FFeq_compact} for mesons and baryons in the case of a rainbow-ladder kernel is shown in Fig.~\ref{fig:FFs_diagrams}.

\subsection{Quark-photon vertex}
If $[K^{(\ell)}]^\mu=0$, then the only additional element needed for the calculation of hadronic currents is the quark-photon vertex $\Gamma^\mu$ (as has been already mentioned, we discuss the coupling of hadrons to electromagnetic fields for concreteness, but the calculation proceeds almost identically for the coupling to other external fields). 
The quark-photon vertex depends on two independent momentum, which can be any combination of the quark momenta $p_1$, $p_2$ and the photon momentum $p_3$. As in the case of bound-states, it is convenient to choose these to be the total photon momentum $Q$ and the relative quark momentum $k$, where
\begin{align}
P = p_3 = p_2 - p_1\;,\qquad k = \left(p_1 + p_2 \right)/2\;,
\end{align}
identical to that of the meson. 

In the DSE/BSE approach the vertex is obtained self-consistently as a solution of an inhomogeneous BSE~\cite{Maris:1999bh}
\begin{align}\label{eqn:inhomogeneousqpvertex}
\Gamma^\mu(k,P) = \Gamma^\mu_0 + \int_q K G_0 \Gamma^\mu(q,P)\;,
\end{align}
where the driving term $\Gamma_0^\mu$ can be taken as the tree-level term $Z_1\gamma^\mu$.

The vertex renormalization $Z_1$ is equal to the quark wave function renormalization $Z_2$ as a consequence of the Abelian Ward-Takahashi identitiy~\cite{Ward:1950xp,Takahashi:1957xn}
\begin{align}\label{eqn:wardtakahashiidentity}
P_\mu \Gamma^\mu(k,P) = S^{-1}(k_+) - S^{-1}(k_-)\;,
\end{align}
where $k_\pm = k\pm P/2$. This can be solved to restrict part of the vertex in terms of the dressing functions of the (inverse) quark propagator~\cite{Ball:1980ay}. Furthermore, under the action of charge conjugation $C$, we have
\begin{align}\label{eqn:qpvertexchargeconjugation}
\overline{\Gamma}^\mu(k,P) = C \Gamma_{\mu}^T(-k,-P)C^T = - \Gamma^\mu(k,-P)\;.
\end{align}
\subsubsection{Covariant Basis}
To encourage consistency in the labelling of the components of the vertex, we follow \cite{Eichmann:2014qva} and similarly employ a common notation
\begin{align}
\Gamma^\mu(k,P) = \Bigl[i\gamma^\mu \Sigma_A + 2 k^\mu\left(i \slashed{k}\Delta_A + \Delta_B\right)\Bigr] + i\sum_{j=1}^8 h_j \tau_j^\mu(k,P)\;.
\end{align}
The first part is that of the Ball-Chiu vertex \cite{Ball:1980ay} and is constrained (in QED) by both the Ward identity and the Ward-Takahashi identity in terms of the quark propagator dressings. The transverse part---right-hand side---on the other hand contains the physical poles and cuts pertaining to e.g.\ the vector meson and higher excitations.

The $\tau_j^\mu$ are functions of the relative quark momenta, $k$, and the incoming gauge boson momentum, $P$. One suitable choice is~\cite{Eichmann:2014qva}
\begin{align}
\begin{array}{rl}
\tau_1^\mu &=t^{\mu\nu}_{PP}\gamma^\nu\;, \\
\tau_5^\mu &=t^{\mu\nu}_{PP} \mathrm{i}k^\nu\;,
\end{array}
\begin{array}{rl}
\tau_2^\mu &= t^{\mu\nu}_{PP} \left(k\cdot P\right) \frac{\mathrm{i}}{2}\left[\gamma^\nu,\slashed{k}\right]\;,\\
\tau_6^\mu &=t^{\mu\nu}_{PP}k^\nu\slashed{k}\;,
\end{array}
\begin{array}{rl}
\tau_3^\mu &=\frac{\mathrm{i}}{2}\left[\gamma^\mu,\slashed{P}\right]\;,\\
\tau_7^\mu &=t^{\mu\nu}_{Pk}\left(k\cdot P\right)\gamma^\nu\;,
\end{array}
\begin{array}{rl}
\tau_4^\mu &=\frac{1}{6}\left[\gamma^\mu,\slashed{k},\slashed{P}\right]\;,\\
\tau_8^\mu &=t^{\mu\nu}_{Pk}\frac{\mathrm{i}}{2}\left[\gamma^\nu,\slashed{k}\right]\;.
\end{array}
\end{align}
where $t^{\mu\nu}_{ab}=\left(a\cdot b\right) \delta^{\mu\nu}-b^\mu a^\nu$. Here, $\tau_1$ and $\tau_5$ correspond to the vector and scalar components of the vertex, respectively, whilst $\tau_3$ and $\tau_4$ in turn relate to the scalar- and vector-anomalous magnetic moments. The triple commutator is defined $[A,B,C]=[A,B]C + [B,C]A + [C,A]B$.

The reason for this particular choice of basis is two-fold. Firstly, it respects the charge conjugation property of the vertex; secondly it is free of kinematic singularities. The dressing functions are subsequently well-behaved, being even in $\left(k\cdot P\right)$ and, in fact, weakly dependent upon the angle. An expansion in e.g. Chebyshev polynomials therefore converges rapidly.

\subsubsection{Integration}
The four-dimensional Euclidean integration over $q$ is performed by recourse to hyperspherical coordinates (see Eq.~\eqref{eq:q3_integration_variable})
%
\begin{align}
\int d^4q = \int_0^\infty dq\, q^3 \int_{-1}^{1} dz\, z^\prime \int_{-1}^{1} dy \int_0^{2\pi} d\phi\;,
\end{align}
with 
$z^\prime= \sqrt{1-z^2}$. The $\phi$ angle may be trivially evaluated since the diagram is planar. Owing to its similarities to the construction and of the meson Bethe-Salpeter equation, see Sec.\ref{sec:mesons}, we tackle the problem in a similar fashion. The only difference is that more caution may be required in evaluating the angular integrals, where  end-point singularities may arise that are best tackled with a tanh-sinh rule. As before, the radial variables are mapped from $[-1,1]$ to $[0,\infty )$. A logarithmic mapping to $[\Lambda_{IR},\Lambda_{UV}]$ with $\Lambda_{IR}\sim 10^{-3}$ and $\Lambda_{UV}\sim 10^{3}$ is sufficiently well-behaved.  

\subsubsection{Results}
\begin{figure}[!ht]
	\begin{center}
		\includegraphics[width=0.98\textwidth]{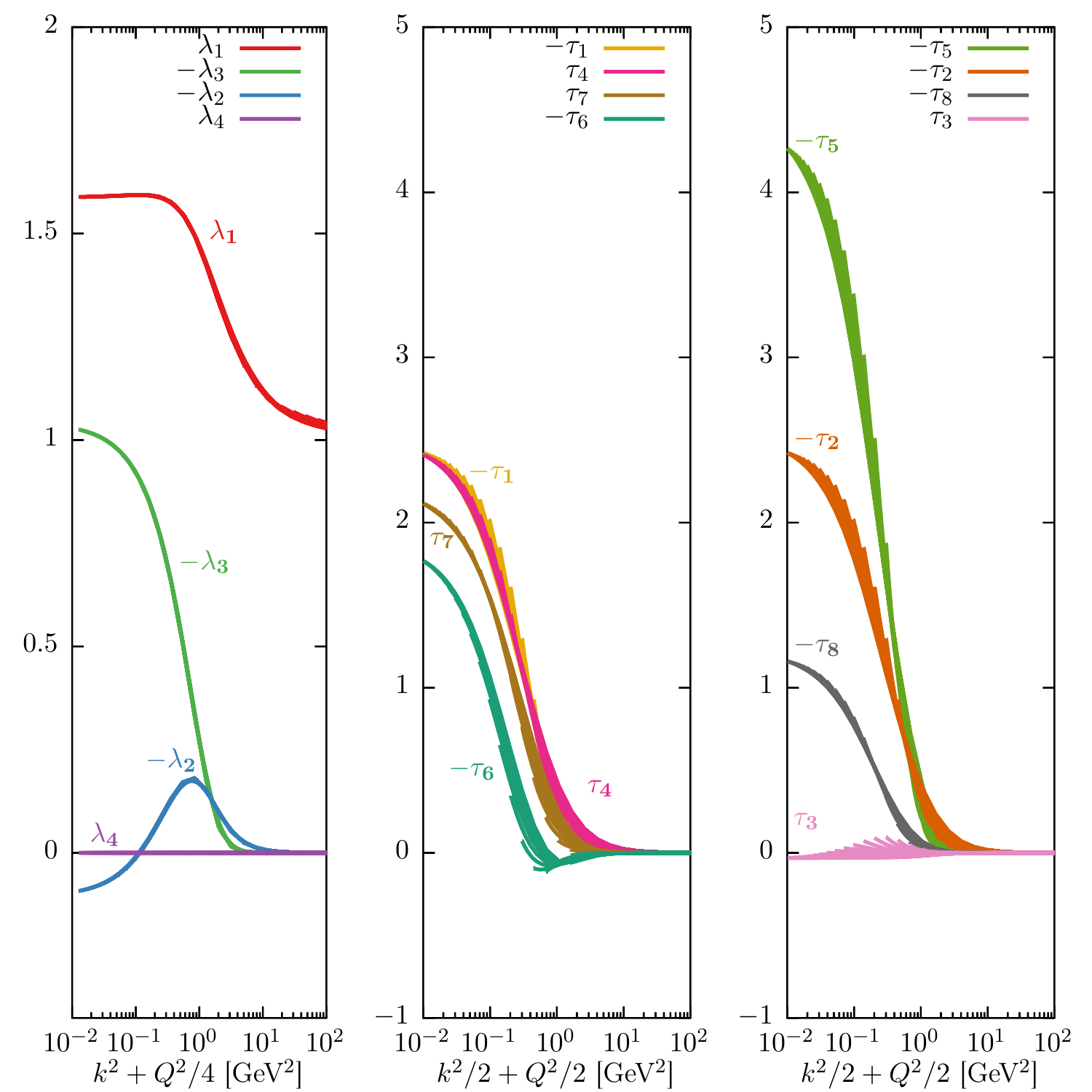}
		\caption{Solution of the quark-photon vertex for a light-quark in the rainbow-ladder approximation, as a function of a single scale $s_0=P^2/4+k^2/3$ in GeV$^2$. Note that we have scaled the dressing functions $\lambda_2$ and $\tau_3$ in order to keep common axes.}\label{fig:solsquark_photon_vertex}
	\end{center}
\end{figure}

We give typical solutions to the quark-photon vertex for light quarks in the rainbow-ladder approximation, using the Maris-Tandy interaction \cite{Maris:1999nt}, in Fig.~\ref{fig:solsquark_photon_vertex}. The dressing functions are parametrized in terms of the $S_3$ permutation group variables $s_0$ and $a,s$ as discussed in Sec.\ref{sec:qgvertex}.

\subsection{Numerical implementation}

Once the quark-photon vertex (or, in general, the gauged kernel) is calculated one can proceed with the numerical implementation of Eq.~\eqref{eq:FFeq_compact}. In the case of the rainbow-ladder truncation this is rather straightforward, in the sense that, with a couple of exceptions discussed below, there are no remarkable optimisations that one can generally use, other than the usual considerations when performing a multi-dimensional integration numerically (see Sec.~\ref{subsubsec:generalities}).

For a meson, the equation to solve is
\begin{align}
J^\mu_{\mathcal{I}'\mathcal{I}}&=
\sum_{\rho\rho';\lambda}\mathcal{Q}^{\rho\rho';\lambda}\int_p 
f^{\rho}_i\bracket{p_f^2,z_f} 
\bar{\tau}^i_{\beta'\alpha'\mathcal{I}'}(p_f,P_f)
\left(
S^{(\lambda_1)}(p_1^f)\Gamma^\mu(p_1,Q)S^{(\lambda_1)}(p_1^i)\right)_{\alpha'\alpha}
\tau^{j}_{\alpha\beta\mathcal{I}}(p_i,P_i)f^{~\rho'}_j\bracket{p^2_i,z_i}
S^{(\lambda_2)}_{\beta\beta'}(k_2)
\nonumber\\&+\textrm{perm.}~, \label{eq:FF_explicit_meson}
\end{align}
with $\tau\bracket{p,P}$ and $f\bracket{p^2,z}$ the covariant meson basis and scalar dressings, respectively.
The charge matrices are defined
\begin{align}
\mathcal{Q}^{\rho\rho';\lambda}=F^{\dagger~\rho}_{ba}\mathds{Q}_{aa'}F^{\rho',\lambda}
_{a'b}~,
\end{align}
where $\mathds{Q}$ is the quark-charge operator.

In practice, one usually works in the so-called Breit frame, where the photon four-momentum is taken as $Q=\bracket{0,0,|Q|,0}$. The initial and final bound states are assumed to be on-shell, which implies that the corresponding total momenta fulfil $P_i^2=-M_i^2$ and $P_f^2=-M_f^2$, with $M_i$ and $M_f$ the initial and final bound-state mass, respectively. This can be achieved by introducing an \emph{average} momentum $P$ defined as
\begin{align}
P=\sqrt{\frac{M^2_{av.}}{\tau}}\bracket{0,0,-\delta,~i\sqrt{\delta^2+\tau\bracket{1+\tau}}}~,\label{eq:average_momentum_FFs}
\end{align}
with
\begin{align}
M^2_{av.}=\frac{M_f^2+M_i^2}{2},\quad 
\delta=\frac{M_f^2-M_i^2}{4M^2_{av.}},\quad
\tau=\frac{Q^2}{4M^2_{av.}}\;.\label{eq:variables_for_FFs}
\end{align}

The boosted momenta in \eqref{eq:FF_explicit_meson} are then defined as
\begin{align}
P_i=P-Q/2~,\quad &P_f=P+Q/2~,\quad &p_i=p-Q/2~,\quad &p_f=p+Q/2~,\\
p_1^i=p_1-Q/2~,\quad &p_1^f=p_1+Q/2~,\quad &p_2^i=p_2+Q/2~,\quad &p_2^f=p_2-Q/2~,
\end{align}
with $p_1$ and $p_2$ defined as in Eq.~\eqref{eq:mesonkinematics} with $p$ the relative momentum.

The most time-consuming part of the four-dimensional integral of Eq.~\eqref{eq:FF_explicit_meson} is, obviously, the evaluation of the integrand. A possible optimisation in this respect is, again, the pre-calculation of the wave function in the rest-frame of the meson
\begin{align}
\omega^{\rho;\lambda_1\lambda_2}_i (p^2,z) \tau^i_{\alpha\beta\mathcal{I}}(p,P)=
S^{(\lambda_1)}_{\alpha\alpha'}(k_1)
\Psi^\rho_{\alpha'\beta'\mathcal{I}}(p,P)
S^{(\lambda_2)}_{\beta'\beta}(k_2)~.
\end{align}
In the integrand, the dressings $\omega^{\rho;\lambda_1\lambda_2}_i (p^2,z)$ need only be interpolated.

For baryons the calculation is a bit more complicated since it involves two different diagrams (see Fig.~\ref{fig:FFs_diagrams}) in addition to their possible permutations. Along the same lines of Sec.~\ref{subsec:three_diagrams_from_one}, the three permuted diagrams can be related to each other \cite{Eichmann:2011vu,Sanchis-Alepuz:2015fcg} and in this case become identical, except for flavour factors.
With the momenta defined as in Eq.~\eqref{eq:defpq}, the simplest diagram is that in which the external field couples to the third quark line. The equation to solve is then
\begin{align}\label{eq:FFeqRL_simple}
J_{\mathcal{I}'\mathcal{I}}^\mu=\sum_{\rho\rho';\lambda}\lsb \mathcal{Q}^{\rho\rho'
}_1(\mathcal{F}_1^{\rho\rho'}J_3^{\rho'\rho'';\lambda}\mathcal{F}_1^{T,
\rho''\rho})_{ \mathcal{I}'\mathcal{I}}^\mu+
\mathcal{Q}^{\rho\rho'}_2(\mathcal{F}_2^{\rho\rho'}J_3^{\rho'\rho'';\lambda}
\mathcal{F}_2^{T,\rho''\rho})_{\mathcal{I}'\mathcal{I}}^\mu+
\mathcal{Q}^{\rho\rho'}_3(J_3^{\rho\rho';\lambda})_{\mathcal{I}'\mathcal{I}}
^\mu\rsb~,
\end{align}
with $\mathcal{F}^{\rho\rho'}_1=F^\rho_{abc}F^{\rho'}_{bca}$, $\mathcal{F}^{\rho\rho'}_2=F^\rho_{abc}F^{\rho'}_{cab}$ and $F$ the flavour amplitudes of the corresponding 
baryons and
\begin{align}
(J_3^{\rho\rho';\lambda})_{\mathcal{I}'\mathcal{I}}^\mu&=\int_p\int_q f_i^\rho \bracket{p_f^2,q_f^2,z_{0,f},z_{1,f},z_{2,f}}\left(f_j^{\rho'}\bracket{p_i^2,q_i^2,z_{0,i},z_{1,i},z_{2,i}}-g_j^{\rho'\lambda}\bracket{p_i^2,q_i^2,z_{0,i},z_{1,i},z_{2,i}}\right) \times\nonumber\\
&~\bar{\tau}
^i_{\beta'\alpha'\mathcal{I}'\gamma'}(p_f,q_f,P_f)\left[S^{(\lambda_1)}_{\alpha'\alpha}(p_1)S^{(\lambda_2)}_{\beta'\beta}
(p_2)\left(S^{(\lambda_3)}(p_3^f)\Gamma^\mu(p_3,Q)S^{(\lambda_3)}(p_3^i)\right)_{\gamma'\gamma}\right]
~\tau^{j}_{\alpha\beta\gamma\mathcal{I}}(p_i,q_i,P_i)~,\label{eq:FF_explicit_baryon}
\end{align}
with
$g_i^{\rho\lambda}$ given by Eq.~\eqref{eq:Faddeev_coeff3}. The boosted momenta are given by
\begin{align}
P_i&=P-Q/2~,\quad &&P_f=P+Q/2~,\\
          p_i&=p-Q/3~,\quad &&p_f=p+Q/3~,\\
          q_i&=q ~,\quad &&q_f = q~,\\
          p_3^i&=p_3-Q/2~,\quad &&p_3^f=p_3+Q/2~,\label{eq:momenta_FF_baryon}
\end{align}
with all other variables defined as in Eqs.~\eqref{eq:defpq} and \eqref{eq:rest_frame_unit_transverse}.

In the baryon case, one obvious optimisation would be not having to evaluate $g_i^{\rho\lambda}$ anew. Therefore, when solving the Faddeev equation it is convenient to store not only the final result of Eq.~\eqref{eq:Faddeev_coeff} but also the partial result in Eq.~\eqref{eq:Faddeev_coeff3}. Additionally, as discussed above, it may be advantageous to pre-calculate the wave function in the rest frame and interpolate its dressings during the integration. 

\subsection{Obstacles in the calculation of hadronic currents}

\subsubsection{Poles of the quark propagator}

As already mentioned in previous sections the quark propagator features complex-conjugate poles in the complex momentum plane. In BSEs, the propagators are probed in a parabolic region of the complex plane, and the location of the poles determines the maximum size of a parabola such that the propagators are analytic inside; equivalently, they determine the minimum value of the apex $-a^2$ of the parabola in the negative real axis. For BSEs, this results in a limitation of the maximum bound-state mass that is accessible. We show below how, when coupling to an external field, this imposes a limit to the momentum transferred to the bound state by the external field.

To be specific, let us consider the situation in baryons, although the reasoning is identical for mesons. As shown in Eq.~\eqref{eq:FF_explicit_baryon}, for the calculation of the current the quark propagators are evaluated at the boosted momenta $p_3^{f/i}=p_3\pm Q/2$. From the definitions \eqref{eq:defpq}, it is clear that in the frame \eqref{eq:average_momentum_FFs} the four-vector $p_3^{f/i}$ can be written as
\begin{align}
p_3^{f/i}=\bracket{r_1,r_2,r_3^\pm,r_4+iI}~,
\end{align}
with all the $r$'s and $I$ real variables. Therefore, the quark propagators are probed in a region given by
\begin{align}
\bracket{p_3^{f/i}}^2=r_1^2+r_2^2+(r_3^\pm)^2+r_4^2-I^2+2ir_4I~,
\end{align}
which is still a parabolic region of the complex plane, with its apex determined by the imaginary part of the average momentum $P$. Therefore, if the quark propagators can be determined down to a minimum apex $-a^2$ before encountering any singularities, the range of allowed values of $Q^2$ can be determined from
\begin{align}
\frac{M^2_{av.}}{9\tau}\bracket{\delta^2+\tau\bracket{1+\tau}}=a^2~,
\end{align}
with $Q^2$ entering via $\tau$. For elastic processes, where  $\delta=0$,
this imposes a maximum limit for $Q^2$ but for inelastic processes, where
$\delta\ne 0$, there is both a maximum and a minimum limit. As an example, for a typical value $a=0.5$, if $M_i=M_f=1$~GeV the maximum photon momentum allowed is $Q^2=5$~GeV$^2$, whereas if $M_i=1$~GeV and $M_f=1.25$~GeV the allowed range for $Q$ is $0.83\le Q^2\le 3.80$~GeV$^2$.

\subsubsection{Non-convergence of the angular Chebyshev expansion}

A different, more technical problem stems from the fact that, when the bound-state BS amplitude is boosted, the variables $z=\widehat{p}\cdot\widehat{P}$ for mesons and $z_0=\widehat{p_T}\cdot\widehat{q_T}$, $z_1=\widehat{p}\cdot\widehat{P}$ and $z_2=\widehat{q}\cdot\widehat{P}$ for baryons become complex and take values outside the complex unit-circle as well. It is easy to see this for the variables $z_1$ and $z_2$ (the angle $z_0$ is more difficult to analyse, although the behaviour is analogous).

\begin{figure}[!ht]
\begin{center}
\hspace*{\fill}
\includegraphics[scale=0.45]{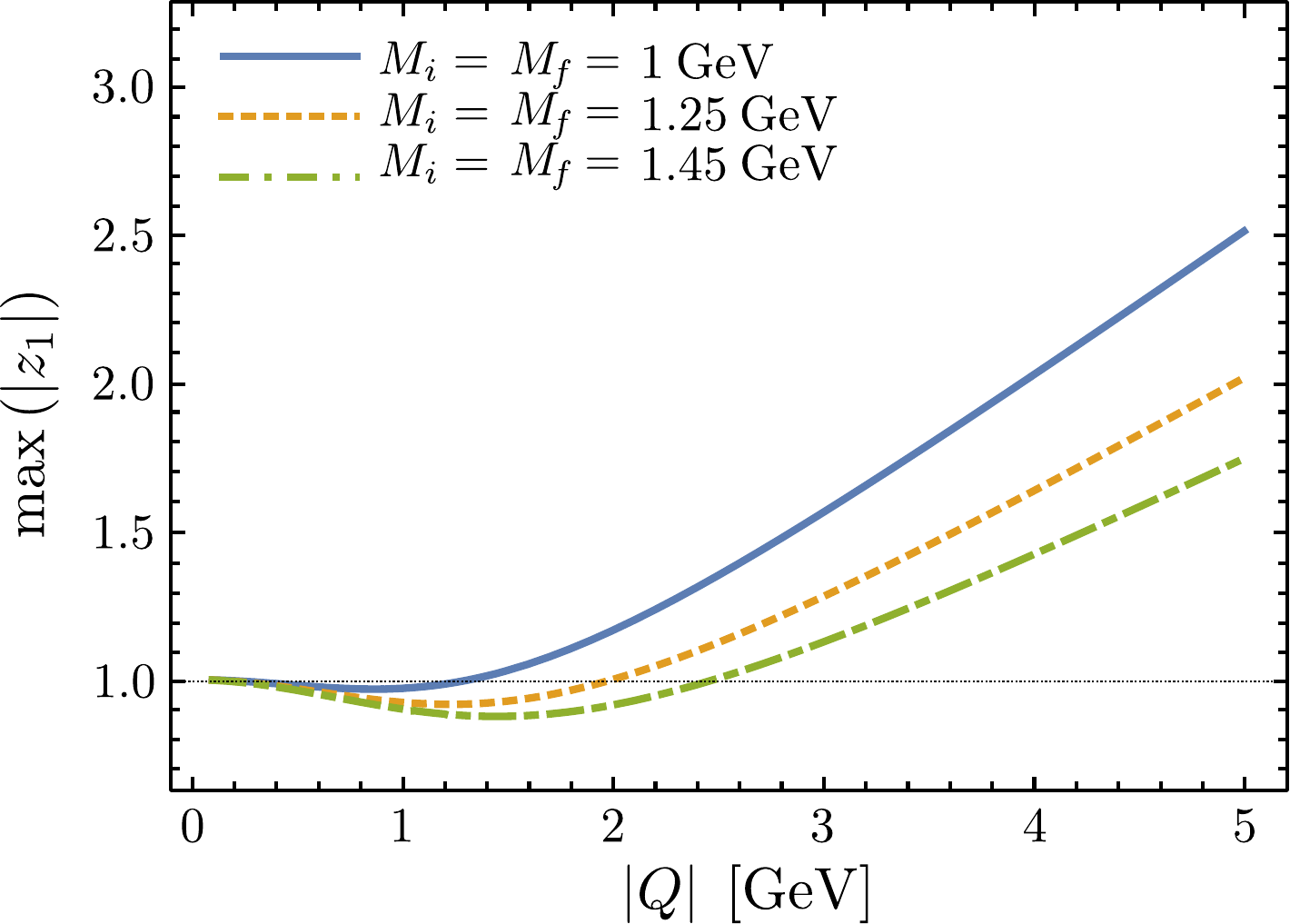}\hfill
\includegraphics[scale=0.45]{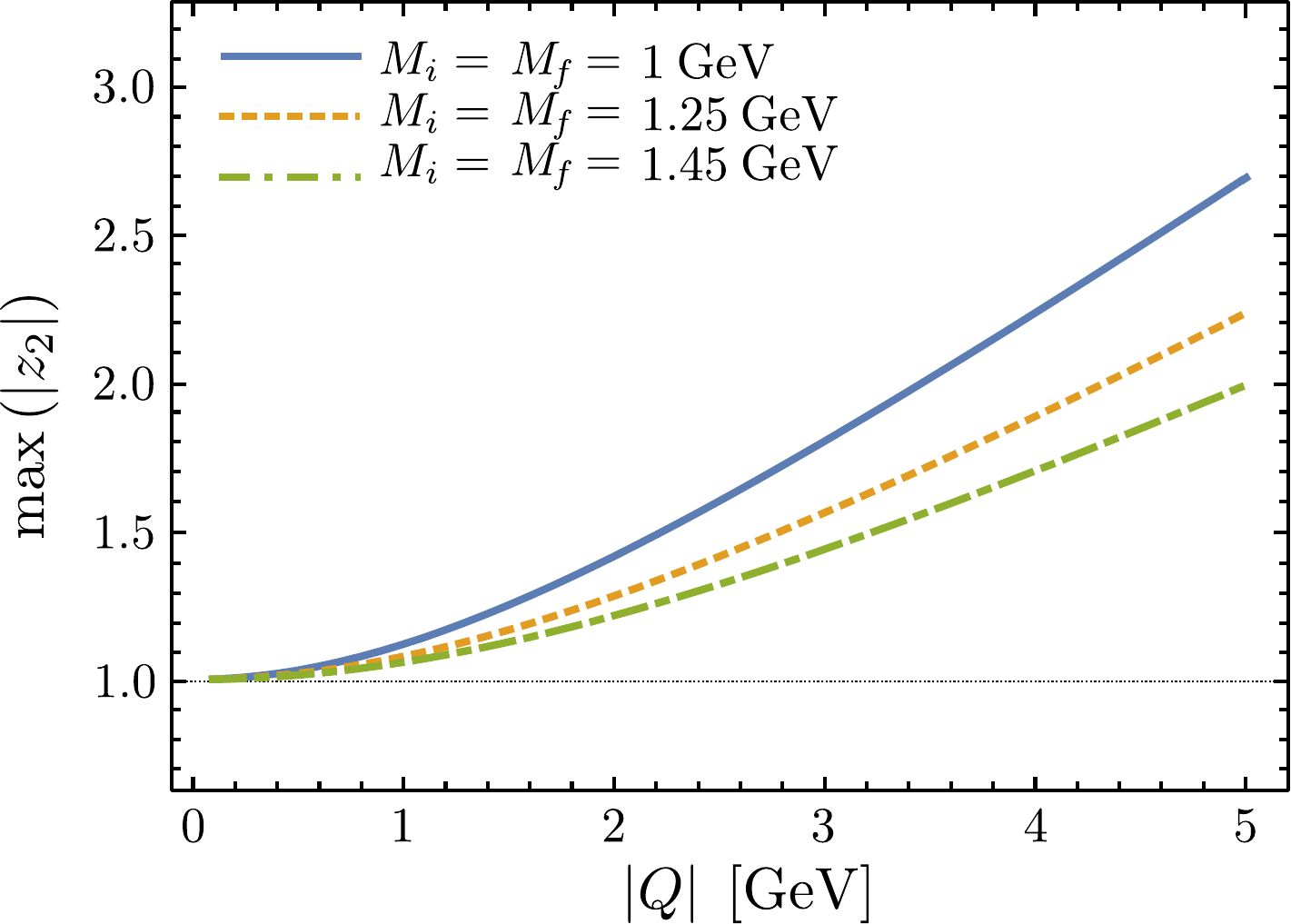}\hspace*{\fill}
\caption{Typical behaviour of the maximum absolute value of the angles $z_1$ and $z_2$, as a function of the transfer momentum $|Q|$, as needed for the calculation of elastic form factors.}\label{fig:angles_elastic_FFs}
\end{center}
\end{figure}
\begin{figure}[!ht]
\begin{center}
\hspace*{\fill}
\includegraphics[scale=0.45]{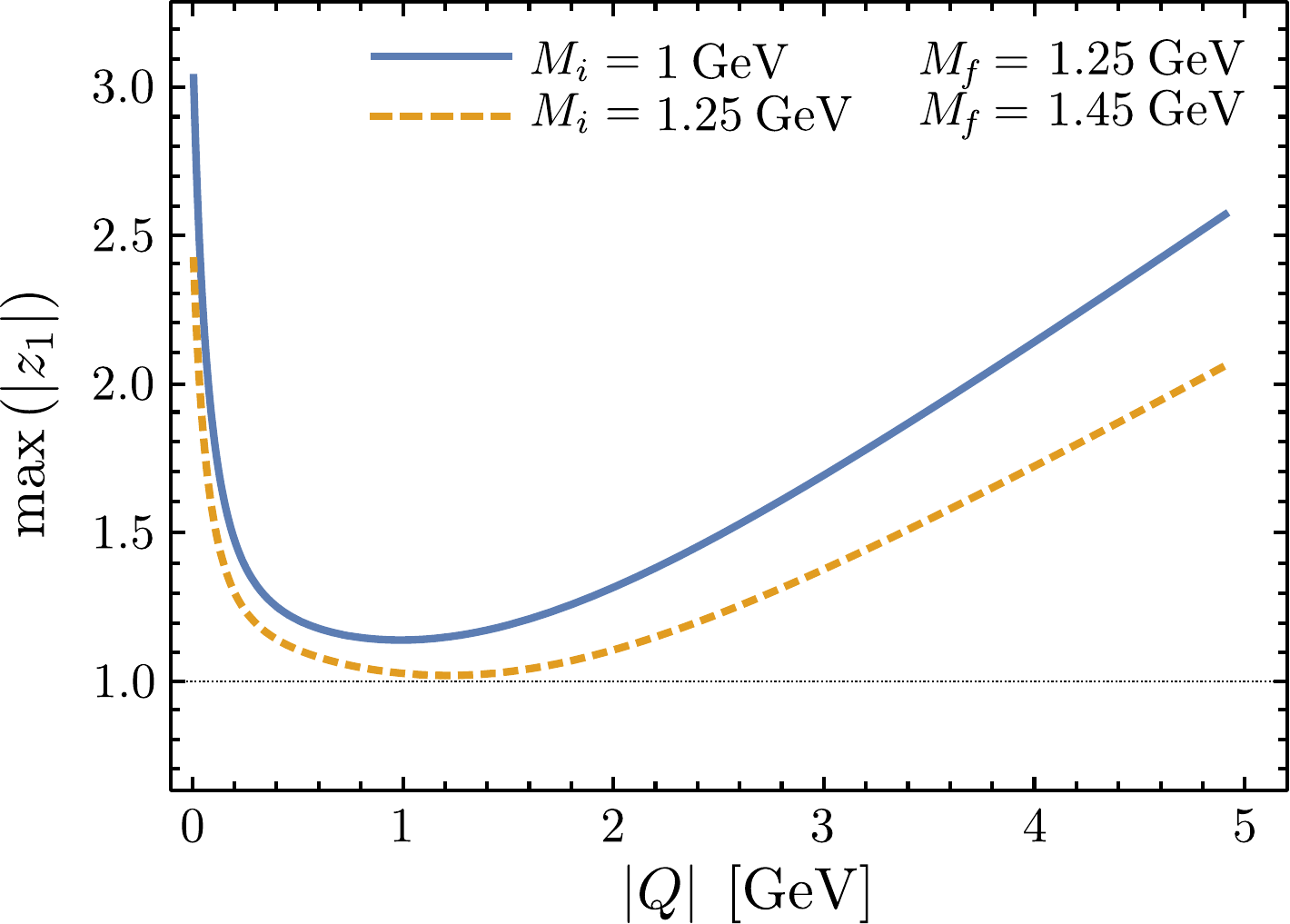}\hfill
\includegraphics[scale=0.45]{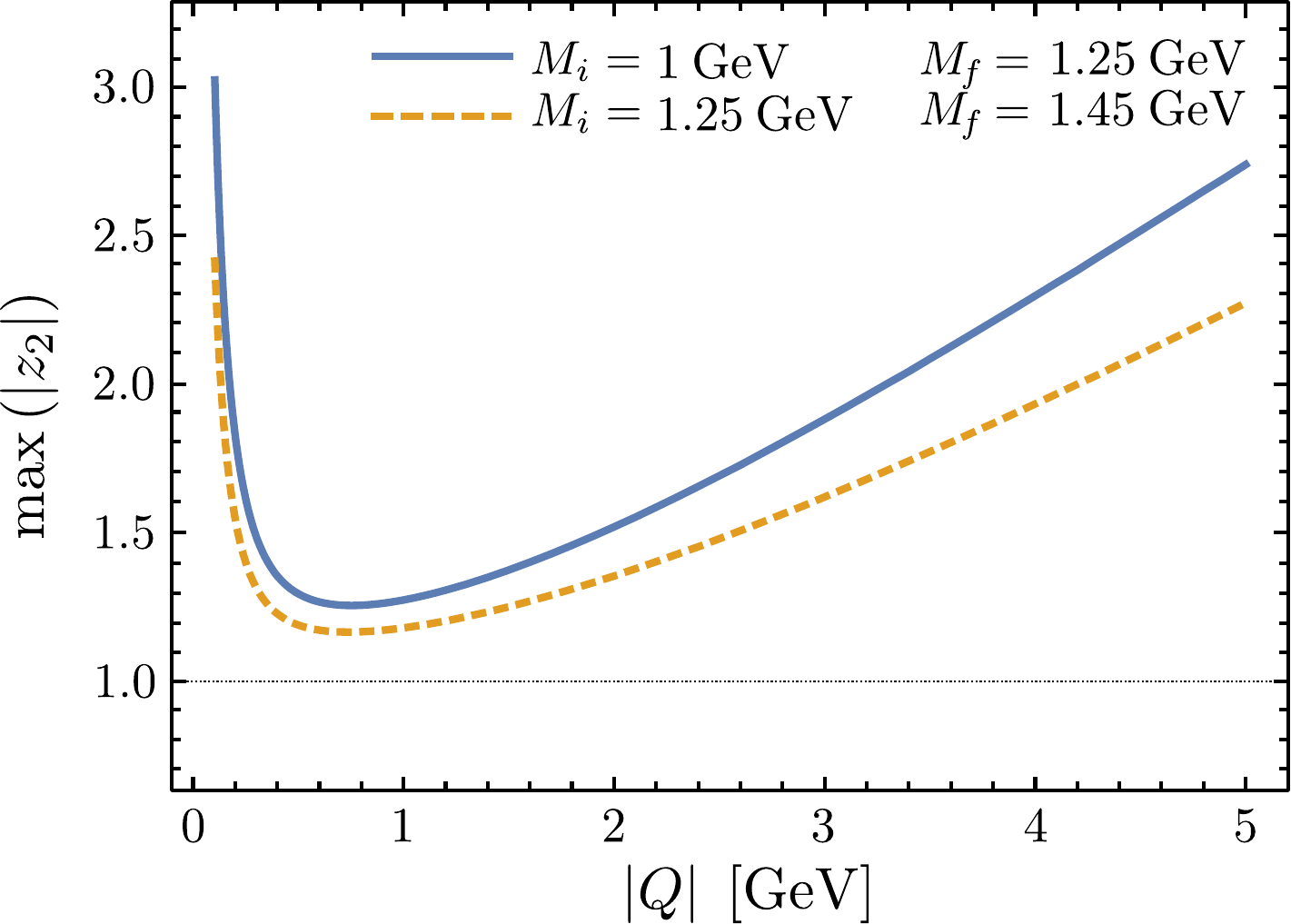}\hspace*{\fill}
\caption{Typical behaviour of the maximum absolute value of the angles $z_1$ and $z_2$, as a function of the transfer momentum $|Q|$,  as needed for the calculation of transition form factors.}\label{fig:angles_transition_FFs}
\end{center}
\end{figure}
From the definitions given in Eqs.~\eqref{eq:average_momentum_FFs} and  \eqref{eq:momenta_FF_baryon} we can write
\begin{align}
\widehat{P_{i/f}}=\frac{1}{iM_{i/f}}\bracket{P\mp Q/2}&\sim\bracket{0,0,-iI,r}~,\nonumber\\
\widehat{p_{i/f}},\widehat{q_{i/f}}&\sim\bracket{r_1,r_2,r_3,r_4}~,\nonumber
\end{align}
where, as before, all the $r$'s and $I$ are real variables. Moreover, from $r_1^2+r_2^2+r_3^2+r_4^2=1$ and $r^2-I^2=1$ we see that, whereas $r_{1\dots 4}\le 1$, this is not the case for $r$ and $I$. The variables $z_{1,2}^{i/f}$ all have the generic form
\begin{align}z_{1,2}^{i/f}\sim& r_4r-ir_3 I = r_4\sqrt{1+I^2}-ir_3I\;,\nonumber\\
\left|z_{1,2}^{i/f}\right|=& r_4^2\bracket{1+I^2}+r_3^2 I^2~.
\end{align}
In the rest-frame of the bound-state we have $I=0$ and clearly these variables behave as regular angles, taking values in the range $\lsb -1,1 \rsb$. When the BS amplitudes are boosted this is no longer the case, as illustrated in Figs.~\ref{fig:angles_elastic_FFs} and \ref{fig:angles_transition_FFs}. The problem appears when the dependence of the BS amplitudes on these variables is expanded in orthogonal polynomials when solving the corresponding BSE, since such an expansion is ill-defined beyond the complex unit-circle and thus cannot be strictly used to interpolate the amplitudes as needed for the calculation of currents (indirectly, the same problem appears if barycentric interpolation is used instead). The only remedy is to introduce additional angles, stemming from scalar products of the type $\widehat{q}\cdot\widehat{Q}$ and $\widehat{p}\cdot\widehat{Q}$ \cite{Maris:2005tt} or, equivalently, expand the angular dependence of the BS amplitudes in terms of the hyperspherical angles directly. This is, however, a significantly more complicated problem to solve numerical especially in the case of beyond rainbow-ladder.

%% file: 6.conclusions.tex
%
\section{Conclusions}
We have presented a comprehensive review of the numerical techniques that we have used to solve Dyson-Schwinger and Bethe-Salpeter equations, with focus on their applications to hadron physics.  
To the best of our knowledge, the most sophisticated techniques developed so far have been described. In particular, the algorithms presented herein allow for the more complex truncations of the DSE/BSE system to be used in practice, thus opening the way for genuine~\emph{ab-initio} calculations of hadron properties in continuum QCD. We hope that this paper enables interested researchers to become practitioners of this challenging and fascinating field.

\section*{Acknowledgements}
We wish to thank the many mentors, colleagues, and students whom we have worked with over the years without whose invaluable input, advice, encouragement, and participation the methodology presented here would not have arisen. We are particularly grateful for interactions with R.~Alkofer, G.~Eichmann, C.~S.~Fischer, W.~Heupel, M.~Huber, A.~Krassnigg, F.~Llanes-Estrada, M.~Mitter, D.~Nicmorus, M.~Pennington, L.~v.~Smekal and C.~Welzbacher.

The work has been supported by the project P29216-N36
from the Austrian Science Fund, FWF, by the Helmholtz International Center 
for FAIR within the LOEWE program of the State of Hesse, and by the 
DFG collaborative research centre TR 16.

%% file: 7.appendix.tex
%
\section{Euclidean space}\label{sec:euclideanspace}
Our frameworks is formulated in Euclidean space with space-time metric $\delta_{\mu\nu}$. We work with hermitian Dirac gamma matrices $\gamma_\mu$, related to the usual Minkowski space matrices via
\begin{align}
\gamma^1 = -i\gamma_M^1\;,\;\;
\gamma^2 = -i\gamma_M^2\;,\;\;
\gamma^3 = -i\gamma_M^3\;,\;\;
\gamma^4 =   \gamma_M^0\;,
\end{align}
and construct fifth Dirac matrix $\gamma_5$ (with $\left(\gamma_5\right)^2=1$) that anti-commutes with the above. One possible choice is
\begin{align}\label{eq:gamma_matrices}
\gamma_1=\begin{pmatrix}
          0 & 0 & 0 & -i \\
          0 & 0 & -i & 0 \\
          0 & i & 0 & 0 \\
          i & 0 & 0 & 0 \\
         \end{pmatrix}~,\quad
&\gamma_2=\begin{pmatrix}
          0 & 0 & 0 & -1 \\
          0 & 0 & 1 & 0 \\
          0 & 1 & 0 & 0 \\
          -1 & 0 & 0 & 0 \\
         \end{pmatrix}~,\quad
&\gamma_3=\begin{pmatrix}
          0 & 0 & -i & 0 \\
          0 & 0 & 0 & i \\
          i & 0 & 0 & 0 \\
          0 & -i & 0 & 0 \\
         \end{pmatrix}~,\quad
&\gamma_4=\begin{pmatrix}
          1 & 0 & 0 & 0 \\
          0 & 1 & 0 & 0 \\
          0 & 0 & -1 & 0 \\
          0 & 0 & 0 & -1 \\
         \end{pmatrix}~,
\end{align}
\begin{align}
\gamma_0=\begin{pmatrix}
          1 & 0 & 0 & 0 \\
          0 & 1 & 0 & 0 \\
          0 & 0 & 1 & 0 \\
          0 & 0 & 0 & 1 \\
         \end{pmatrix}~,\quad
&\gamma_5=\begin{pmatrix}
          0 & 0 & 1 & 0 \\
          0 & 0 & 0 & 1 \\
          1 & 0 & 0 & 0 \\
          0 & 1 & 0 & 0 \\
         \end{pmatrix}~.
\end{align}
The usual Clifford algebra then applies
\begin{align}
\left\{\gamma_\mu,\gamma_\nu \right\} = 2\delta_{\mu\nu}\;.
\end{align}
For bound-state calculations we will need the charge conjugation matrix. In terms of the above gamma matrices, it is defined by $C=\gamma_4\gamma_2$, with its inverse given by $\mathcal{C}^{-1}=\mathcal{C}^T = -\mathcal{C}$

\section{Numerical methods}\label{sec:numerics}
\subsection{Chebyshev expansion}\label{sec:chebyshev}
It is convenient in many situations in DSE and BSE calculations to expand the momentum dependence of dressing functions in orthogonal polynomials. Particularly useful in this respect are Chebyshev polynomials. For $x\in [-1,1]$, the polynomials of the first kind, defined by the recurrence relation
\begin{align}
T_n(x)=2 x T_{n-1}(x)-T_{n-2}(x)~,\qquad n=2,3,\dots~,
\end{align}
with $T_0(x)=1/\sqrt{2}$ and $T_1(x)=x$, the following discrete orthogonality relation is exact (see, e.g. \cite{mason2002chebyshev})
\begin{align}
\sum_{k=1}^{N}T_i(x_k)T_j(x_k)=a_{ij}~,
\end{align}
where $x_k=\cos\theta_k$, with $\theta_k=(k-\nicefrac{1}{2})~\pi/N$, are the zeroes of $T_{N}(x)$ and $a_{ii}=N/2$ for $i\le N-1$, $a_{ij}=0$ for $i\neq j$ and $0\le i,j \le N-1$.
If a function $f(x)$ is expanded as
\begin{align}
f(x)=\sum_{i=0}^{n} c_i T_i(x)~,
\end{align}
the coefficients of the expansion are given by
\begin{align}
c_i=\frac{2}{n+1}\sum_{i=1}^{n+1}f(x_k)T_i(x_k)~.
\end{align}

The polynomials of the second kind are defined by the recurrence relation
\begin{align}
U_n(x)=2 x U_{n-1}(x)-U_{n-2}(x)~,\qquad N=2,3,\dots~,
\end{align}
with $U_0(x)=1$ and $U_1(x)=2 x$. The weighted polynomials $\sqrt{1-x^2}U_i(x)$ obey the following exact discrete orthogonality relation
\begin{align}
\sum_{k=1}^{N}(1-y_k^2)U_i(y_k)U_j(y_k)=b_{ij}~,
\end{align}
with $y_k=\cos k\pi/(N+1)$ the zeroes of $\sqrt{1-x^2}U_{N}(x)$ and $b_{ii}=(N+1)/2$ for $i\le N-1$, $b_{ij}=0$ for $i\neq j$ and and $0\le i,j \le N-1$. If a function $g(x)$ is expanded as
\begin{align}
g(x)=\sum_{i=0}^{n} c_i U_i(x)~,
\end{align}
the coefficients of the expansion are given by
\begin{align}
c_i=\frac{2}{n+2}\sum_{i=1}^{n+1}(1-y_k^2)g(y_k)U_i(y_k)\;.
\end{align}
\subsection{Barycentric Lagrange interpolation}\label{sec:barycentric}
In general, the polynomial $p$ that interpolates a function $f$ at the points $x_j$ has a unique solution and can be written in Lagrange form
\begin{align}
p(x) = \sum_{j=0}^n f_j\; l_j(x)\;,\qquad l_j(x) =\prod_{k=0,k\ne j}^n \frac{x-x_k}{x_j-x_k}\;.
\end{align}
where $l_j(x)$ is the Lagrange polynomial. As it stands, each of evaluation of $p(x)$ is of complexity $\mathcal{O}(n^2)$ and is not stable numerically.
It can be rewritten in terms of barycentric weights~\cite{doi:10.1137/S0036144502417715,Wang:2014}
\begin{align}
l_j(x) &= l(x) \frac{w_j}{x-x_j}\;,\qquad \textrm{with}\;\;
l(x) = \prod_{k=0}^n \left(x-x_k\right)\;,\qquad \textrm{and}\;\;
w_j  = \frac{1}{l^\prime(x)} = \frac{1}{\prod_{k=0,k\ne j}^n\left(x_j-x_k\right)}\;.
\end{align}
Then the first form of the barycentric interpolation formula is
\begin{align}
p(x) = l(x)\sum_{j=0}^n \frac{w_j}{x-x_j} f_j\;.
\end{align}
The next step is to note that
\begin{align}
1 = \sum_{j=0}^n l_j(x) = l(x) \sum_{j=0}^n \frac{w_j}{x-x_j}\;,
\end{align}
and hence one may write the second form of the barycentric interpolation formula
%
\begin{align}\label{eqn:barycentricformula}
p(x) =\left.\sum_{j=0}^n \frac{w_j}{x-x_j}f_j\;\; \middle/ \;\; \sum_{j=0}^n \frac{w_j}{x-x_j} \right.\;.
\end{align}

Note the similarity with the Cauchy integration formula of in Eq.~\ref{eqn:cauchystable}.

\section{Solving the Eigenvalue Problem}\label{sec:eigenvalues}
We previously stated that Bethe-Salpeter equation is solved as an eigenvalue equation of the form
\begin{align}
\mathbf{A}\,\mathbf{x} = \lambda\mathbf{x}\;,
\end{align}
where solutions $\lambda(P^2)=1$ for discrete $P^2=-M^2_i$ correspond to bound-states. The ground state is identified with $\lambda_0$, the largest eigenvalue and hence its eigenvector can be extracted by the method of power iteration
\begin{align}
\mathbf{x}_{i+1} = \mathbf{A}\,\mathbf{x}_i\;,\qquad
\lambda[\mathbf{x}_i] = \frac{\mathbf{x}_i^\dag \mathbf{A} \mathbf{x}_i}{\mathbf{x}_i^\dag \mathbf{x}_i}
= \frac{\mathbf{x}_{i+1}^\dag \mathbf{x}_i}{\mathbf{x}_i^\dag \mathbf{x}_i} \;,
\end{align}
where $\lambda$ is estimated from the Rayleigh Quotient and $x_i^\dag$ is the conjugate transpose of $x_i$. Convergence is typically judged from the residual at the $k$-th iterative step $\mathbf{R}^k = \lambda\mathbf{x}^k-\mathbf{A}\mathbf{x}^k$
by demanding that its \emph{magnitude} is below some threshold.

To access excited states we need to employ more robust techniques such as Arnoldi factorisation: see ARPACK~\cite{doi:10.1137/S0895479895281484,doi:10.1137/1.9780898719628}.

\section{Basis decomposition}\label{sec:basis}
\subsection{Symmetric trace-free tensors}\label{sec:symmetrictracefree}
The process of incorporating total spin $J$ into both our mesons and baryons involves a common starting point, the construction of symmetric and trace-free tensors~\cite{Zemach:1968zz}.

Denoting symmetrisation by parenthesis around the indices, and targeting the indices $a_1,\ldots,a_n$ yields
\begin{align}\label{eqn:symmetrictensor}
S^{a_1 a_2 \ldots a_n}
=
R^{(a_1 a_2\ldots a_n)}
=
\frac{1}{n!}\sum_{\sigma\in \mathscr{S}_n} R^{a_{\sigma(1)}a_{\sigma(2)}\ldots a_{\sigma(n)}}\;,
\end{align}
where $\mathscr{S}_n$ is the symmetric group on the set $\left\{1,2,\ldots, n \right\}$. From this, we construct a tensor that is trace-free in any pair of indices i.e. $S^{\kappa\kappa a_3\ldots a_n}=0$. Thus
\begin{align}\label{eqn:symmetrictracefreetensor}
T^{a_1a_2\ldots a_n }
&=
\sum_{i=0}^{\lfloor n/2\rfloor}
\beta_i^n\;
\delta_T^{(a_1 a_2}\cdots \delta_T^{a_{2i-1}a_{2i}}
S^{a_{2i+1}\ldots a_n )\kappa_1\ldots\kappa_i\kappa_1\ldots\kappa_i}\;.
\end{align}
Here $\lfloor \cdot \rfloor$ is the floor function. The coefficients $\beta$ are given by
\begin{align}
\beta_i^n = (-1)^i \frac{n!(2n-2i-1)!!}{(n-2i)!(2n-1)!!(2i)!!}\;,
\end{align}
where $n!! = \prod_{k=0}^{\lceil n/2\rceil -1}(n-2k)$ is the double factorial with $\lceil \cdot \rceil$ the ceiling function.

Note the symmetrisation over $n$ indices which is accompanied by a $1/n!$. If we work with unique permutations, i.e. take into consideration the symmetry properties of the kronecker delta and the uncontracted indices of $S$ then we collect an additional combinatorial factor of $2^j j! (n-2j)! /n!$. This comes from the number of different ways that $n$ indices can be split into $j$ (ordered) pairs and the symmetry under permutations of the $n-2j$ free indices of $S$. Then the modified coefficients reduce to
\begin{align}
\tilde{\beta}_i^n = \left(-1\right)^i  \prod_{j=0}^{i-1}\left(2n-2j-1\right)^{-1}\;.
\end{align}

Let us give some examples. For convenience we compress the indices by writing $a_1 a_2\ldots a_n=12\ldots n$:
\begin{align}
T^{1}                  =&& S^{1 }
& \;,\\
T^{1 2 }          =&& S^{1 2 }
&- \frac{1}{3} \delta_T^{1 2} S^{\kappa\kappa }\;,\\
T^{1 2 3 } =&& S^{1 2 3 }
&- \frac{1}{5} \left( \delta_T^{1 2} S^{3\kappa\kappa }
+ \delta_T^{2 3} S^{1\kappa\kappa }
+ \delta_T^{3 1} S^{2\kappa\kappa }\right) \;,\\
T^{1 2 3 4 } =&& S^{1 2 3 4 }
&- \frac{1}{7}\left(
\delta_T^{1 2} S^{3 4\kappa\kappa }
+ \delta_T^{1 3} S^{2 4\kappa\kappa }
+ \delta_T^{1 4} S^{2 3\kappa\kappa }
+ \delta_T^{2 3} S^{1 4\kappa\kappa }
+ \delta_T^{2 4} S^{1 3\kappa\kappa }
+ \delta_T^{3 4} S^{1 2\kappa\kappa }
\right)\nonumber \\
&& &+ \frac{1}{35}
\left( \delta_T^{1 2}\delta_T^{3 4}
+\delta_T^{1 3}\delta_T^{2 4}
+\delta_T^{1 4}\delta_T^{2 3}\right) S^{\kappa_1\kappa_2\kappa_1\kappa_2 }\;,\\
T^{1 2 3 4 5 } =&& S^{1 2 3 4 5 }
&- \frac{1}{9}\Big(
\delta_T^{1 2} S^{3 4 5\kappa\kappa }
+ \delta_T^{1 3} S^{2 4 5\kappa\kappa }
+ \delta_T^{1 4} S^{2 3 5\kappa\kappa }
+ \delta_T^{1 5} S^{2 3 4\kappa\kappa }
+ \delta_T^{2 3} S^{1 4 5\kappa\kappa }
\nonumber\\
&& & + \phantom{\frac{1}{9}\Big(}
\delta_T^{2 4} S^{1 3 5\kappa\kappa }
+ \delta_T^{2 5} S^{1 3 4\kappa\kappa }
+ \delta_T^{3 4} S^{1 2 5\kappa\kappa }
+ \delta_T^{3 5} S^{1 2 4\kappa\kappa }
+ \delta_T^{4 5} S^{1 2 3\kappa\kappa }
\Big)\nonumber \\
&& &+ \frac{1}{63}\bigg[
\left( \delta_T^{1 2}\delta_T^{3 4}
+\delta_T^{1 3}\delta_T^{2 4}
+\delta_T^{1 4}\delta_T^{2 3}\right)
S^{5\kappa_1\kappa_2\kappa_1\kappa_2 }
+
\left( \delta_T^{2 5}\delta_T^{3 4}
+\delta_T^{3 5}\delta_T^{2 4}
+\delta_T^{4 5}\delta_T^{2 3}\right)
S^{1\kappa_1\kappa_2\kappa_1\kappa_2 }\nonumber\\
&& &+\phantom{\frac{1}{63}\Big(}
\left( \delta_T^{1 5}\delta_T^{3 4}
+\delta_T^{1 3}\delta_T^{4 5}
+\delta_T^{1 4}\delta_T^{3 5}\right)
S^{2\kappa_1\kappa_2\kappa_1\kappa_2 }
+
\left( \delta_T^{1 2}\delta_T^{4 5}
+\delta_T^{1 5}\delta_T^{2 4}
+\delta_T^{1 4}\delta_T^{2 5}\right)
S^{3\kappa_1\kappa_2\kappa_1\kappa_2 }\nonumber\\
&& &+\phantom{\frac{1}{63}\Big(}
\left( \delta_T^{1 2}\delta_T^{3 5}
+\delta_T^{1 3}\delta_T^{2 5}
+\delta_T^{1 5}\delta_T^{2 3}\right)
S^{4\kappa_1\kappa_2\kappa_1\kappa_2 }
\bigg]\;.
\end{align}
\subsection{Mesons}\label{sec:mesonbasis}
The angular momentum tensors for bound state of two quarks with total spin $J$ are constructed from
\begin{align}
R_1^{\mu_1\mu_2\ldots\mu_J} = \widehat{p_T}^{\mu_1}\widehat{p_T}^{\mu_2}\cdots \widehat{p_T}^{\mu_J}
\;,\;\;\;\;
R_2^{\mu_1\mu_2\ldots\mu_J} = \gamma^{\mu_1}_\perp \widehat{p_T}^{\mu_2}\cdots \widehat{p_T}^{\mu_J}\;,
\end{align}
by imposing symmetrisation and tracelessness as discussed above; for a more detailed discussion see Refs.~\cite{LlewellynSmith:1969az,Krassnigg:2010mh,Fischer:2014xha}.

Our basic covariants are then given by
\begin{align}
G_1 =  \mathds{1}\;,\;\;\;\;
G_2 =  \widehat{\slashed{p}_T}\;,
\end{align}
for (pseudo)scalar mesons, while
\begin{align}
G_1^{\mu_1\ldots\mu_J} = T_1^{\mu_1\ldots\mu_J} \; \mathds{1}\;,\;\;\;
G_2^{\mu_1\ldots\mu_J} = T_1^{\mu_1\ldots\mu_J} \; \widehat{\slashed{p}_T}\;,\;\;\;
G_3^{\mu_1\ldots\mu_J} = T_2^{\mu_1\ldots\mu_J} \; \mathds{1}\;,\;\;\;
G_4^{\mu_1\ldots\mu_J} = T_2^{\mu_1\ldots\mu_J} \; \widehat{\slashed{p}_T}\;,
\end{align}
for those with total spin one or greater. Here $\Lambda_\pm = \frac{1}{2}\left(\mathds{1}\pm\slashed{\hat{P}}\right)$ must be folded in to give four and eight tensor structures, respectively. Organising these covariants according partial waves we have
\begin{center}
	\begin{tabular}{ccl}
		\multicolumn{3}{c}{$J=0$} \\
		\toprule
		s & l &  element \\
		\midrule
		$1$   &  $1$     &   $G_2$  \\
		$0$   &  $0$     &   $G_1$ \\
		&& \\
		&& \\
		\bottomrule
	\end{tabular}
\hspace{0.5cm}
\begin{tabular}{ccl}
	\multicolumn{3}{c}{$J=1$} \\
	\toprule
	s & l &  element \\
	\midrule
	$1$   &  $2$   &   $3G_2^{\mu_1} - G_3^{\mu_1}$  \\
	$1$   &  $1$   &   $ G_4^{\mu_1} - G_1^{\mu_1}$  \\
	$1$   &  $0$   &   $ G_3^{\mu_1}$                \\
	$0$   &  $1$   &   $ G_1^{\mu_1}$                \\
	\bottomrule
\end{tabular}
\hspace{0.5cm}
\begin{tabular}{ccl}
	\multicolumn{3}{c}{$J>1$} \\
	\toprule
	s & l &  element \\
	\midrule
	$1$   &  $J+1$   &   $G_3^{\mu_1\ldots\mu_J} - \frac{J+1}{J} G_2^{\mu_1\ldots\mu_J}$ \\
	$1$   &  $J$     &   $G_4^{\mu_1\ldots\mu_J}$       \\
	$1$   &  $J-1$   &   $G_2^{\mu_1\ldots\mu_J} + G_3^{\mu_1\ldots\mu_J}$ \\
	$0$   &  $J$     &   $G_1^{\mu_1\ldots\mu_J}$       \\
	\bottomrule
\end{tabular}
\end{center}
To determine the partial wave decomposition, we exploited the total spin operator $S^2$
\begin{align}
\left[\mathbf{S}^2\right]_{ab,cd}
=
\frac{3}{2} \left[\mathds{1}\right]_{ab}  \left[\mathds{1}\right]_{cd} -
\left[\gamma^\mu_T \gamma_5 \slashed{\hat{P}}\right]_{ab}
\left[\slashed{\hat{P}}\gamma_5 \gamma^\mu_T\right]_{cd}\;,
\end{align}
that acts upon the basis elements as
\begin{align}
\left[\mathbf{S}^2\right]_{ab,cd} \left[G_i^{\mu_1\mu_2\ldots\mu_J}\right]_{bc}\;.
\end{align}
Similarly, for quark angular momentum we use the differential operator $\mathbf{L}^2$
\begin{align}
\mathbf{L}^2 = \widehat{p_T}^\mu \widehat{p_T}^\nu \partial_\mu \partial_\nu
+ 2\widehat{p_T}^\mu  \partial_\mu
- \widehat{p_T}^2 \partial^2\;,
\end{align}
where we write $\widehat{p_T}^\mu = \left(x_1, x_2, x_3, 0\right)$ and take $x_1=x_2=0$, $x_3=1$ at the end.

\subsection{Baryons}\label{sec:baryonbasis}
A basic set of Dirac covariants for the baryon is provided by 64 elements~\cite{Henriques:1975uh,Carimalo:1992ia}
\begin{align}
S_{ij}(\widehat{p_T},\widehat{q_t},\hat{P}) &=
\phantom{\gamma_5}\Gamma_i(\widehat{p_T},\widehat{q_t}) \Lambda_\pm(\hat{P})\gamma_5 C
\otimes
\phantom{\gamma_5}\Gamma_j(\widehat{p_T},\widehat{q_t})\Lambda_+(\hat{P}) \;,\\
P_{ij}(\widehat{p_T},\widehat{q_t},\hat{P}) &=
         \gamma_5 \Gamma_i(\widehat{p_T},\widehat{q_t}) \Lambda_\pm(\hat{P}) \gamma_5 C
\otimes
         \gamma_5 \Gamma_j(\widehat{p_T},\widehat{q_t})\Lambda_+(\hat{P}) \;,
\end{align}
where $\Lambda_\pm(\hat{P})=(\mathds{1}\pm\widehat{\slashed{P}})/2$ and $i,j=1,\ldots,4$ spans the elements
\begin{align}
\Gamma_i\bracket{\widehat{p_T},\widehat{q_t}}=&
\left\{
\mathds{1},\;
\frac{1}{2}\left[\widehat{\slashed{p}_T}, \widehat{\slashed{q}_t} \right],\;
\widehat{\slashed{p}_T},\;
\widehat{\slashed{q}_t} \right\}\;.
\end{align}

The basis for spin $j=n+\nicefrac{1}{2}$ is constructed by incorporating contractions of the Rarita-Schwinger projector with $n$ momenta
\begin{align}
S_{ij}^{pq\cdots q} = \left(S_{ij}(\widehat{p_T},\widehat{q_t},\hat{P})\right) \left( \mathds{1} \otimes p^{\mu_1}q^{\mu_2}\cdots q^{\mu_n} \mathds{P}^{\mu_1\cdots\mu_j;\nu_1\cdots\nu_j}\right)\;.
\end{align}
For the Rarita-Schwinger projector for arbitrary spin, we use
\begin{align}
\mathds{P}^{\mu_1\cdots\mu_n;\nu_1\cdots\nu_n} = \left(\frac{n+1}{2n+3}\right)\gamma^\mu_T\gamma_T^\nu P^{\mu\mu_1\cdots\mu_n;\nu\nu_1\cdots\nu_n}\;,
\end{align}
where $P^{\mu\mu_1\cdots\mu_n;\nu\nu_1\cdots\nu_n}$ is symmetric trace-free raw tensor (discussed in Sec.\ref{sec:symmetrictracefree}) composed from the $n+1$ products of the metric tensor.

One choice for the basis of the Nucleon baryon arranged by partial waves is as given in Table~\ref{tab:nucleonpwdbasis}. We similarly list the partial wave basis for the Delta baryon in Table~\ref{tab:deltapwdbasis}.

\begin{table}[ht!]
\begin{align*}
\begin{array}{c}
\begin{pmatrix*}[r]
0 & \frac{1}{2} & P_{33}+P_{44}+S_{22} \\
0 & \frac{1}{2} & P_{22}+S_{33}+S_{44} \\
0 & \frac{1}{2} & P_{11} \\
0 & \frac{1}{2} & S_{11} \\
\end{pmatrix*} \\
\vspace{1.25cm}\\
\begin{pmatrix*}[r]
2 & \frac{3}{2} & P_{44}-S_{22} \\
2 & \frac{3}{2} & P_{42}+S_{24} \\
2 & \frac{3}{2} & 2 (P_{34}+P_{43}) \\
2 & \frac{3}{2} & P_{33}-\frac{1}{2} P_{44}-\frac{1}{2} S_{22} \\
2 & \frac{3}{2} & 2 (P_{32}+S_{23}) \\
2 & \frac{3}{2} & P_{23}+S_{32} \\
2 & \frac{3}{2} & P_{22}-S_{33} \\
2 & \frac{3}{2} & -\frac{1}{2} P_{22}-\frac{1}{2} S_{33}+S_{44} \\
2 & \frac{3}{2} & 2 (S_{34}+S_{43}) \\
2 & \frac{3}{2} & 2 (P_{24}+S_{42}) \\
\end{pmatrix*}
\end{array}
\qquad
\begin{pmatrix*}[r]
1 & \frac{3}{2} & 2 (P_{34}-P_{43}+2 S_{21}) \\
1 & \frac{3}{2} & 2 (P_{32}-2 P_{41}-S_{23}) \\
1 & \frac{3}{2} & \frac{1}{2} (2 P_{31}+P_{42}-S_{24}) \\
1 & \frac{3}{2} & -2 (2 P_{21}+S_{34}-S_{43}) \\
1 & \frac{3}{2} & -2 P_{24}+4 S_{31}+2 S_{42} \\
1 & \frac{3}{2} & \frac{1}{2} (P_{23}-S_{32})+S_{41} \\
1 & \frac{1}{2} & -P_{34}+P_{43}+S_{21} \\
1 & \frac{1}{2} & -P_{31}+P_{42}-S_{24} \\
1 & \frac{1}{2} & P_{32}+P_{41}-S_{23} \\
1 & \frac{1}{2} & P_{24}+S_{31}-S_{42} \\
1 & \frac{1}{2} & P_{23}-S_{32}-S_{41} \\
1 & \frac{1}{2} & P_{21}-S_{34}+S_{43} \\
1 & \frac{1}{2} & P_{14} \\
1 & \frac{1}{2} & P_{13} \\
1 & \frac{1}{2} & P_{12} \\
1 & \frac{1}{2} & S_{14} \\
1 & \frac{1}{2} & S_{13} \\
1 & \frac{1}{2} & S_{12} \\
\end{pmatrix*}
\end{align*}
\caption{The tensor basis for the Nucleon baryon arranged by partial waves. The first column corresponds to  the quark orbital angular momentum $l$ while the second column refers to the total quark spin $s$. See the text for details of the symbolic notation used for the basis elements given in the third column.\label{tab:nucleonpwdbasis}}
\end{table}


\begin{table}[ht!]
	\begin{align*}
	\begin{array}{c}
	\begin{pmatrix*}[r]
	0 & \frac{3}{2} & P^p_{31}+P^q_{41}-S^p_{24}+S^q_{23} \\
	0 & \frac{3}{2} & 2 (-P^p_{24}+P^q_{23}+S^p_{31}+S^q_{41})
	\end{pmatrix*} \\
	\vspace{1.25cm}\\
	\begin{pmatrix*}[r]
	2 & \frac{3}{2} & \frac{1}{2} (P^p_{33}-2 P^p_{44}+2 P^q_{34}+P^q_{43}+S^p_{22}+S^q_{21}) \\
	2 & \frac{3}{2} & \frac{1}{2} (-P^p_{34}-2 P^p_{43}+2 P^q_{33}-P^q_{44}+S^p_{21}-S^q_{22}) \\
	2 & \frac{3}{2} & \frac{1}{2} (P^p_{31}-P^p_{42}+2 P^q_{32}-P^q_{41}-2 S^p_{24}+S^q_{23}) \\
	2 & \frac{3}{2} & \frac{1}{2} (-P^p_{32}+2 P^p_{41}+2 P^q_{31}+P^q_{42}+S^p_{23}-S^q_{24}) \\
	2 & \frac{3}{2} & P^p_{31}+P^p_{42}-P^q_{41}-S^q_{23} \\
	2 & \frac{3}{2} & \frac{1}{2} (P^p_{24}+P^q_{23}-2 S^p_{31}-S^p_{42}-S^q_{32})+S^q_{41} \\
	2 & \frac{3}{2} & -\frac{2}{3} (P^p_{22}+P^q_{21}+S^p_{33}-2 S^p_{44}+2 S^q_{34}+S^q_{43}) \\
	2 & \frac{3}{2} & -\frac{2}{3} (P^p_{21}-P^q_{22}-S^p_{34}-2 S^p_{43}+2 S^q_{33}-S^q_{44}) \\
	2 & \frac{3}{2} & P^p_{24}-P^q_{23}+S^p_{42}-S^q_{32} \\
	2 & \frac{3}{2} & \frac{2}{3} (P^p_{23}-P^q_{24}-S^p_{32}+2 (S^p_{41}+S^q_{31})+S^q_{42}) \\
	2 & \frac{1}{2} & -\frac{1}{2} P^p_{33}-\frac{1}{2} P^p_{44}-P^q_{34}+P^q_{43}-\frac{1}{2} S^p_{22}+S^q_{21} \\
	2 & \frac{1}{2} & \frac{1}{2} (P^p_{31}-P^p_{42}+2 P^q_{32}+2 P^q_{41}+S^p_{24}-2 S^q_{23}) \\
	2 & \frac{1}{2} & -\frac{2}{3} (2 P^p_{34}-2 P^p_{43}-P^q_{33}-P^q_{44}-2 S^p_{21}-S^q_{22}) \\
	2 & \frac{1}{2} & -P^p_{31}+P^p_{42}-S^p_{24} \\
	2 & \frac{1}{2} & \frac{2}{3} (P^p_{32}+P^p_{41}+P^q_{31}-P^q_{42}-S^p_{23}+S^q_{24}) \\
	2 & \frac{1}{2} & \frac{1}{2} (-P^p_{24}+2 P^q_{23}-S^p_{31}+S^p_{42}-2 (S^q_{32}+S^q_{41})) \\
	2 & \frac{1}{2} & -\frac{1}{2} P^p_{22}+P^q_{21}-\frac{1}{2} S^p_{33}-\frac{1}{2} S^p_{44}-S^q_{34}+S^q_{43} \\
	2 & \frac{1}{2} & P^p_{24}+S^p_{31}-S^p_{42} \\
	2 & \frac{1}{2} & \frac{2}{3} (P^p_{23}-P^q_{24}-S^p_{32}-S^p_{41}-S^q_{31}+S^q_{42}) \\
	2 & \frac{1}{2} & \frac{2}{3} (2 P^p_{21}+P^q_{22}-2 S^p_{34}+2 S^p_{43}+S^q_{33}+S^q_{44}) \\
	2 & \frac{1}{2} & \frac{1}{2} P^p_{13}+P^q_{14} \\
	2 & \frac{1}{2} & \frac{1}{2} P^p_{11}+P^q_{12} \\
	2 & \frac{1}{2} & \frac{2}{3} (P^p_{14}+P^q_{13}) \\
	2 & \frac{1}{2} & P^p_{13} \\
	2 & \frac{1}{2} & \frac{2}{3} (2 P^p_{12}-P^q_{11}) \\
	2 & \frac{1}{2} & \frac{1}{2} S^p_{13}+S^q_{14} \\
	2 & \frac{1}{2} & \frac{1}{2} S^p_{11}+S^q_{12} \\
	2 & \frac{1}{2} & \frac{2}{3} (S^p_{14}+S^q_{13}) \\
	2 & \frac{1}{2} & S^p_{13} \\
	2 & \frac{1}{2} & \frac{2}{3} (2 S^p_{12}-S^q_{11})
	\end{pmatrix*}
	\end{array}
	\qquad
	\begin{array}{c}
	\begin{pmatrix*}[r]
	1 & \frac{3}{2} & \frac{1}{2} (3 P^p_{34}-2 P^q_{33}+P^q_{44}+3 S^p_{21}+S^q_{22}) \\
	1 & \frac{3}{2} & P^p_{33}-2 P^p_{44}+3 P^q_{43}+S^p_{22}-3 S^q_{21} \\
	1 & \frac{3}{2} & \frac{1}{2} (P^p_{32}-2 P^p_{41}+2 P^q_{31}+P^q_{42}-S^p_{23}-S^q_{24}) \\
	1 & \frac{3}{2} & -2 (P^p_{22}-3 P^q_{21}+S^p_{33}-2 S^p_{44}+3 S^q_{43}) \\
	1 & \frac{3}{2} & -2 (3 P^p_{21}+P^q_{22}+3 S^p_{34}-2 S^q_{33}+S^q_{44}) \\
	1 & \frac{3}{2} & \frac{2}{5} (P^p_{23}+P^q_{24}-S^p_{32}+2 S^p_{41}-2 S^q_{31}-S^q_{42}) \\
	1 & \frac{1}{2} & P^p_{33}+P^p_{44}+S^p_{22} \\
	1 & \frac{1}{2} & -2 (P^q_{33}+P^q_{44}+S^q_{22}) \\
	1 & \frac{1}{2} & 2 (P^p_{32}+P^p_{41}-P^q_{31}+P^q_{42}-S^p_{23}-S^q_{24}) \\
	1 & \frac{1}{2} & 2 (P^p_{23}+P^q_{24}-S^p_{32}-S^p_{41}+S^q_{31}-S^q_{42}) \\
	1 & \frac{1}{2} & P^p_{22}+S^p_{33}+S^p_{44} \\
	1 & \frac{1}{2} & -2 (P^q_{22}+S^q_{33}+S^q_{44}) \\
	1 & \frac{1}{2} & 2 (P^p_{14}-P^q_{13}) \\
	1 & \frac{1}{2} & 2 P^q_{11} \\
	1 & \frac{1}{2} & P^p_{11} \\
	1 & \frac{1}{2} & 2 (S^p_{14}-S^q_{13}) \\
	1 & \frac{1}{2} & 2 S^q_{11} \\
	1 & \frac{1}{2} & S^p_{11}
	\end{pmatrix*}\\
	\vspace{1.25cm}\\
	\begin{pmatrix*}[r]
	3 & \frac{3}{2} & \frac{1}{2} (P^p_{43}+2 P^q_{44}-S^p_{21}-2 S^q_{22}) \\
	3 & \frac{3}{2} & \frac{1}{2} (P^p_{41}+2 P^q_{42}+S^p_{23})+S^q_{24} \\
	3 & \frac{3}{2} & \frac{1}{2} (P^p_{33}+P^p_{44}+2 P^q_{34}+P^q_{43}-2 S^p_{22}+S^q_{21}) \\
	3 & \frac{3}{2} & \frac{1}{2} (P^p_{31}+2 P^p_{42}+2 P^q_{32}-P^q_{41}+S^p_{24}+S^q_{23}) \\
	3 & \frac{3}{2} & \frac{1}{6} (4 P^p_{34}+5 P^p_{43}+4 P^q_{33}-2 P^q_{44}-S^p_{21}-2 S^q_{22}) \\
	3 & \frac{3}{2} & \frac{1}{3} (3 P^p_{33}-P^p_{44}-P^q_{43}-2 S^p_{22}+S^q_{21}) \\
	3 & \frac{3}{2} & \frac{1}{6} (8 P^p_{32}-P^p_{41}-4 P^q_{31}-2 P^q_{42}+7 S^p_{23}+2 S^q_{24}) \\
	3 & \frac{3}{2} & \frac{1}{6} (7 P^p_{23}+2 P^q_{24}+8 S^p_{32}-S^p_{41}-4 S^q_{31}-2 S^q_{42}) \\
	3 & \frac{3}{2} & \frac{1}{12} (14 P^p_{22}-7 P^q_{21}-16 S^p_{33}+2 S^p_{44}-5 S^q_{34}+2 S^q_{43}) \\
	3 & \frac{3}{2} & \frac{1}{16} (-7 P^p_{21}-14 P^q_{22}-2 S^p_{34}+5 S^p_{43}-2 S^q_{33})+S^q_{44} \\
	3 & \frac{3}{2} & \frac{1}{2} (P^p_{23}+2 P^q_{24}+S^p_{41})+S^q_{42} \\
	3 & \frac{3}{2} & \frac{1}{3} (-2 P^p_{22}+P^q_{21}+2 S^p_{44}+3 S^q_{34}+2 S^q_{43}) \\
	3 & \frac{3}{2} & \frac{1}{3} (-P^p_{21}-2 P^q_{22}+2 S^p_{34}+3 S^p_{43}+2 S^q_{33}) \\
	3 & \frac{3}{2} & \frac{2}{3} (P^p_{24}+P^q_{23}+S^p_{31}+2 (S^p_{42}+S^q_{32})-S^q_{41}) \\
	\end{pmatrix*}
	\end{array}
	\end{align*}
\caption{The tensor basis for the Delta baryon arranged by partial waves. The first column corresponds to the quark orbital angular momentum $l$ while the second column refers to the total quark spin $s$. See the text for details of the symbolic notation used for the basis elements given in the third column.\label{tab:deltapwdbasis}}
\end{table}